\documentclass{article}

\usepackage{amsmath, amsthm, amssymb}
\usepackage{graphicx}
\usepackage{verbatim}
\usepackage{natbib}
\usepackage{caption}
\usepackage{subcaption}
\usepackage{fancyvrb}
\usepackage{enumerate}
\usepackage{relsize}
\usepackage{algorithm}
\usepackage{hyperref}
\usepackage{algpseudocode}
\usepackage{enumitem}
\usepackage{mathtools}
\usepackage{multirow}
\usepackage{booktabs}
\usepackage[most]{tcolorbox}

\usepackage{hyperref}
\usepackage[margin=1.5in]{geometry}
\hypersetup{colorlinks,citecolor=blue,urlcolor=blue,linkcolor=blue}

\usepackage{notations}

\theoremstyle{plain}
\newtheorem{proposition}{Proposition}

\newtheorem{corollary}[proposition]{Corollary}
\newtheorem{lemma}[proposition]{Lemma}
\newtheorem{theorem}[proposition]{Theorem}

\theoremstyle{definition}
\newtheorem{example}{Example}
\newtheorem{definition}{Definition}
\newtheorem{assumption}{Assumption}

\theoremstyle{remark}

\newtheorem{remark}{Remark}

\algrenewcommand{\algorithmicrequire}{\textbf{Input:}}
\algrenewcommand{\algorithmicensure}{\textbf{Output:}}

\title{Spectral Design of Random-Duration Switchbacks}
\author{Yuchen Hu\\Columbia University}

\date{Draft version \ifcase\month\or
January\or February\or March\or April\or May\or June\or
July\or August\or September\or October\or November\or December\fi \ \number%
\year\ \  }

\begin{document}
\maketitle

\begin{abstract}
A switchback experiment alternates an entire system between treatment and control over time. It is especially useful when interactions between units can undermine standard unit-level experiments.
Switchback experiments are commonly implemented on fixed temporal grids, which impose highly structured restrictions on when treatment can switch. We study a broader class of random-duration switchbacks, in which treatment and control alternate across runs whose durations are drawn from a common distribution. Under finite carryover, we show that the mean squared error of the Horvitz-Thompson estimator has a simple frequency-domain representation governed by how temporal outcome patterns align with design-induced imbalance and contamination patterns. This representation yields a tractable worst-case design criterion that can be computed directly from the duration distribution. Optimizing over even a simple two-rate family produces a run-age-dependent switching rule that reduces the asymptotic worst-case mean squared error by at least 32\% relative to the standard independently randomized fixed-block switchback.
\end{abstract}

\section{Introduction}
\label{sec:intro}

Switchback experiments are widely used for online experimentation on large-scale platforms, especially marketplaces where unit-level randomization may create substantial interference \citep{chamandy2016experimentation,kastelman2018switchback,kohavi2020trustworthy}. By exposing the entire system to one treatment at a time, switchbacks avoid distortions that can arise when interacting units are simultaneously assigned to different treatments. They can also be useful when only a single experimental unit is available, or when large differences across units make repeated comparisons within the same unit more informative \citep{bojinov2019time,liang2025randomization}. 

A common way to run a switchback experiment is to divide time into blocks of the same length and alternate treatment and control across consecutive blocks \citep{chamandy2016experimentation,kastelman2018switchback}. This design is simple and keeps treatment and control closely balanced, but it also creates a highly regular treatment pattern. Because outcomes often follow strong daily or weekly cycles, this regularity can line up systematically with those cycles. For example, if a ridesharing platform switches treatment every hour, then the 8--9am period is assigned to the same treatment arm every day, so a systematic morning peak during this hour can be mistaken for a treatment effect. One practical workaround is to use a block length that does not align with obvious cycles,
but any fixed block length still creates a regular temporal pattern. Shared operational rhythms or other experiments using similar schedules can generate variation at the same frequency, and relevant periodicities may not be known in advance \citep{xiong2023bias}.

A natural way to break this regularity is to assign each fixed-length block independently to treatment or control with probability one half \citep{kastelman2018switchback,bojinov2020design,hu2022switchback}. This removes the deterministic alternation between treatment arms, but the switching rule remains highly structured: the chance of switching is zero within each block and jumps to one half at every block boundary. After merging consecutive blocks with the same assignment, each treatment run lasts for a geometric number of blocks, a by-product of independent block-level randomization on a fixed temporal grid. Despite its widespread use, however, there is no obvious reason why the resulting distribution of run durations should be optimal in general.

In this paper, we take the durations of the realized treatment runs as the primitive design choice. After each switch, the duration of the next treatment run is drawn from a distribution \(G\), and the treatment label alternates when the run ends. This gives a simple random-duration construction that contains both alternating fixed blocks and independently randomized fixed blocks as special cases, while allowing the experimenter to control the frequency and regularity of actual switches more directly. We then ask how the duration distribution \(G\) should be chosen to minimize worst-case estimation error under temporal carryover.
In taking this worst-case view, we allow the environment to be nonstationary and vary arbitrarily over time.

We show that, when expressed in the frequency domain, the estimation error decomposes into two spectral-overlap terms: one between common-outcome variation and design-induced imbalance, and the other between treatment-effect variation and design-induced contamination.
Thus, a design performs poorly when either imbalance or contamination has a large spectral peak at a frequency where the corresponding outcome component also varies strongly. Intuitively, a poor design can systematically mistake baseline variation for a treatment effect, or leave periods with relatively high or low treatment effects disproportionately contaminated. This representation immediately yields a tractable worst-case proxy that can be computed directly from the duration distribution $G$ through its characteristic and truncated characteristic functions. 

Optimizing over even a simple two-rate family of random-duration designs reduces the asymptotic worst-case mean squared error by at least \(32\%\) relative to the standard switchback benchmark. 
One particularly simple implementation of this design is to draw a new treatment duration from an exponential distribution with a low switching rate at each switch. If this duration would end within the period over which outcomes may still be affected by the previous treatment, it is used as drawn. Otherwise, the current treatment is kept in place through this carryover period, after which its remaining duration is drawn from a second exponential distribution with a much higher switching rate.
Interestingly, the optimal early switching rate remains strictly positive, so the design occasionally switches before reaching uncontaminated exposure. 
Altogether, these results suggest that substantial gains can come from allowing the switching rate to depend on the age of the current treatment run.

\subsection{Related work}
\label{sec:literature}

Switchback experiments have a long history dating back to early agricultural experiments \citep{brandt1938tests,cochran1941double}. More recently, they have been widely adopted for online experimentation in marketplaces and other settings where unit-level randomization may lead to substantial interference \citep{chamandy2016experimentation, kastelman2018switchback, kohavi2020trustworthy}. 
A central design consideration in these experiments is the tradeoff between switching frequently to maintain treatment--control balance over time and keeping treatment in place long enough for the effects of previous assignments to dissipate.
Existing approaches thus differ in how they model this temporal dependence.
\citet{bojinov2020design} take a design-based approach under a finite carryover horizon and derive a minimax-optimal randomization schedule, whereas \citet{glynn2020adaptive} and \citet{hu2022switchback} model the underlying system dynamically using Markovian structures. 
More recent work has also explored model-assisted and data-adaptive approaches that use additional features of the outcome process to guide switchback design and analysis \citep{xiong2023bias, chen2025efficient, wen2025unraveling, ni2025enhancing}.

More broadly, switchback experiments belong to the literature on temporal experiments in which treatments are repeatedly assigned to the same evolving unit. 
This repeated-treatment setting arises in time-series experiments and \(N\)-of-1 designs, where a single unit is observed under a sequence of treatment assignments \citep{bojinov2019time,basse2023minimax,liang2025randomization,guo2026experimental}, as well as in micro-randomized trials, which repeatedly randomize treatment to learn how treatment effects depend on the evolving history \citep{liao2016sample,hu2023offpolicy}. In many applications, the unit also has an evolving state that carries information about the effect of past treatments. This has motivated methods that explicitly exploit the underlying dynamics \citep{farias2022markovian,shi2023dynamic,li2024experimenting,hu2025targeting,johari2026estimation}.

Across these approaches, the treatment assignment process is typically parameterized through predetermined temporal partitions such as block lengths or randomization points. 
Recent work has begun to relax this restriction. For example, \citet{xiong2023bias} allow the boundaries of treatment intervals to be randomized using prior information, while other approaches allow treatment assignments to depend on the realized experimental history through sequential rerandomization or Markovian switching rules \citep{wager2026causal,zeng2026sequential,guo2026experimental}. Such history dependence is nevertheless typically local, for example through the previous treatment or a low-dimensional summary of recent observations. We instead take the distribution of realized treatment durations as the primitive design choice and allow the propensity to switch to depend directly on the age of the current run. This makes it possible to control not only how often treatment switches, but also the regularity and tail behavior of the resulting switch times.

Carryover effects can be viewed as a form of interference across time, where treatment assigned at one time may affect outcomes observed later. This connects switchback experiments to the broader literature on experimental design under interference. 
A classical approach to interference is cluster randomization, which assigns groups of interacting units jointly \citep{ugander2013graph,leung2022rate}. 
Of particular relevance, recent work has also moved beyond the use of a fixed clustering, either by randomizing the partition to average over unfavorable cluster boundaries and improve the probability of observing informative exposure patterns \citep{ugander2023randomized,ni2023design}, or by constructing assignments that make exposure conditions relevant to the target estimand sufficiently likely \citep{cai2024independent,kandiros2024conflict}.

Our assignment process is an alternating renewal process, a classical two-state process in which the system alternates between states for random durations \citep{cox1962renewal,ross1996stochastic}. 
The frequency-domain properties of renewal-driven two-state processes have also been studied in the literature on random telegraph signals and renewal noise \citep{lowen1993fractal,neuts1989square}. 
However, this literature often takes the renewal law as given and uses its spectrum to characterize the resulting stochastic process; in our setting, the renewal law is instead the experimental design variable. 
More broadly, frequency-domain methods have a long history in experimental design for dynamic system identification, where the input spectrum is optimized to learn an underlying dynamical model \citep{zarrop1979optimal,rojas2007robust}. In our setting, the spectrum is instead induced by the randomization process, and we optimize the underlying duration distribution using a model-free worst-case criterion.

\section{Random-duration switchback designs}
\label{sec:renewal}

Rather than describing a switchback through a fixed grid of possible switching
times, we describe it directly through its realized treatment runs. Let
$\mathcal G$ denote the collection of distributions on $(0,\infty)$ with finite
mean. For $G\in\mathcal G$, 
we consider experimental designs in which successive treatment runs have durations governed by
$G$, and the treatment label switches whenever a run ends.

\begin{definition}[Random-duration switchback]
\label{defi:renewal}
For $G \in \mathcal G$, let $D_G$ be the design whose switch times form a
stationary renewal process with inter-switch distribution $G$.
Given the switch times, the run containing time \(0\) is assigned to treatment or control with equal probability, and treatment labels alternate across successive runs.
\end{definition}

To implement the random-duration switchback, the experimenter first chooses the treatment at time \(0\) by a fair coin. 
Following the standard equilibrium initialization of a renewal process \citep[e.g., Chapter 3 of][]{ross1996stochastic}, to initialize the design, draw the duration \(R_0\) of the run containing time \(0\) from the length-biased distribution associated with \(G\), and place time \(0\) uniformly within that run. At the end of the current run, switch the treatment label, draw a new duration independently from \(G\), and repeat this procedure throughout the experiment.

With the random-duration design, the experimenter's design choice is the duration distribution $G$. It determines how frequently treatment switches, as well as how regular or variable the resulting switch times are. Different choices of $G$ therefore
allow the experimenter to control both the typical duration of treatment runs
and the randomness of the switching pattern.

\subsection{Fixed-block designs as special cases}
\label{sec:special_case}

The random-duration formulation makes clear that the two fixed-block designs commonly employed correspond to highly structured choices of the duration distribution $G$. In both cases, durations are restricted to multiples of a fixed block length $\ell$, so treatment can switch only at a predetermined set of run ages. The two designs differ only in what happens at these eligible switching times. For both designs, we randomize the block-grid origin uniformly over \([0,\ell)\) to obtain their stationary versions.

\begin{example}[Alternating fixed blocks] 
\label{exam:fixed}
Divide time into blocks of length $\ell$, randomize the initial treatment label, and alternate treatment and control at
every block boundary. This is $D_G$ with
\[
    G = \delta_\ell,
\]
the point mass at $\ell$.
\end{example}

\begin{example}[Regular switchback]\footnote{\cite{bojinov2020design} use the term regular switchback for a broader class that allows unequal spacing between prespecified randomization points. Their minimax-optimal design, however, uses equally
spaced randomization points away from the experiment boundaries. We use the
term here for this equal-block implementation, which is also common in the
temporal experiment literature \citep{hu2022switchback,xiong2023bias,wen2025unraveling}.}
\label{exam:regular}
Divide time into blocks of length $\ell$ and assign each block independently to
treatment or control with probability one half. After merging adjacent blocks
with the same label, this is $D_G$ with
\[
    G = \sum_{k=1}^\infty 2^{-k}\delta_{k\ell}.
\]
Equivalently, a treatment run consists of a geometric number of blocks,
$R=\ell K$ with $K\sim\operatorname{Geom}(1/2)$ on $\{1,2,\ldots\}$.
\end{example}

Although both designs belong to the random-duration class, they impose strong additional
restrictions on when treatment is allowed to switch. In both cases, switches are confined
to a predetermined lattice of run ages. To see this more explicitly, consider for these lattice-valued designs
the conditional switching probability
\[
    q_G(r) := \PP[G]{R=r \mid R\ge r}.
\]
For alternating fixed blocks, $q_G(\ell)=1$, with no possibility of switching at
any other run age. For a regular switchback, $q_G(k\ell)=\frac12, \, k=1,2,\ldots$,
and the switching probability is zero away from the lattice
$\{\ell,2\ell,\ldots\}$. 

\begin{figure}[t]
\centering
\includegraphics[width=\textwidth]{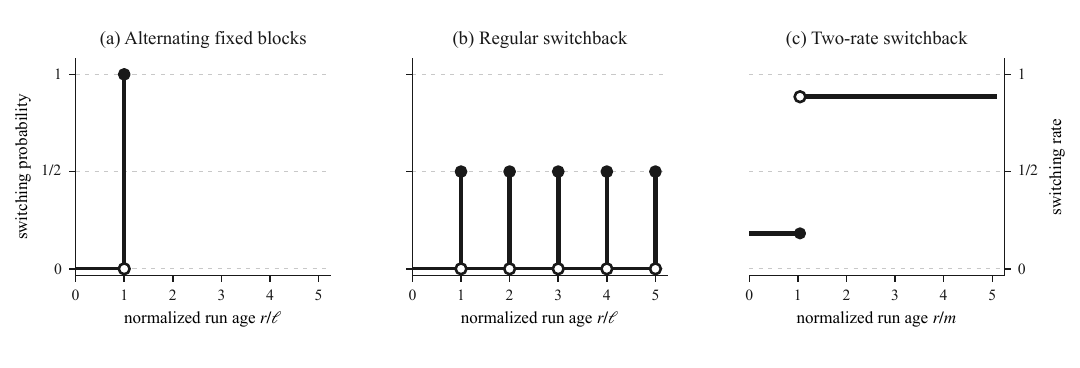}
\caption{Switching behavior as a function of run age under three switchback designs. Panels (a)-(b) show conditional switching probabilities for the two fixed-block designs, where switching is impossible between the lattice ages. Panel (c) shows an illustrative two-rate random-duration switchback, where the continuous-time switching rate changes with run age and is not tied to a predetermined lattice. Run age is normalized by the block length \(\ell\) in panels (a)–(b) and by the carryover horizon \(m\) in panel (c).}
\label{fig:switching_rules}
\end{figure}

We illustrate these restrictions in Figure~\ref{fig:switching_rules}. 
The two conventional constructions impose two quite different---and somewhat
incidental---duration structures. 
Alternating fixed blocks make every run
exactly the same duration, which creates a perfectly regular switching pattern that
can be exploited by periodic adversarial outcome patterns. 
A regular switchback
breaks this regularity by randomizing each block independently, but this
mechanically induces geometric durations, with potentially large variation in
how long a treatment persists. 
In both cases, the duration distribution is a
by-product of the block-level construction and can be tuned only indirectly
through the block length $\ell$. 

In contrast, a random-duration design removes this lattice restriction completely: switching can depend directly on the age of the current run, as illustrated by the simple age-dependent switching rule in the right panel. As we will see shortly, this additional flexibility can matter substantially. Even a simple two-rate piecewise-exponential design yields a markedly better worst-case guarantee than the fixed-block designs above.

\section{The minimax design problem}

We now turn to the choice of the duration distribution $G$. Our goal is to
design the switchback to estimate the effect of global treatment as accurately
as possible while guarding against unfavorable temporal outcome patterns. We
formalize this objective through a worst-case mean squared error criterion.

For a treatment assignment path $\mathbf w:\mathbb R\to\{0,1\}$, let $Y_t(\mathbf w)$ denote the potential
outcome at time $t$ under $\mathbf w$. For $a\in\{0,1\}$, let $\mathbf a$
denote the constant assignment path that assigns $a$ throughout. Our estimand is
the time-averaged effect of global treatment relative to global control,
\begin{equation}
    \tau_T
    :=
    \frac{1}{T}
    \int_0^T
    \{
    Y_t(\mathbf 1)-Y_t(\mathbf 0)\}\,dt.
\label{eq:GATE}
\end{equation}
This is the continuous-time analogue of the global average treatment effect
\citep{ugander2013graph,bojinov2020design,xiong2023bias,hu2022switchback}.
Throughout, we take a design-based perspective in which the potential outcomes are fixed and may vary arbitrarily over time \citep{imbens2015causal}.

To identify and estimate this global treatment effect from a switchback experiment, we impose
the standard non-anticipation condition for time-series potential outcomes
\citep{bojinov2019time}, together with a finite carryover horizon as commonly
assumed in switchback experiments \citep{bojinov2020design}. Under this assumption, the outcome at time $t$ depends only on the treatment assignment during the preceding $m$ units of time. In particular, if treatment $a$ has been maintained
throughout $[t-m,t]$, the observed outcome at time $t$ equals $Y_t(\mathbf a)$.

\begin{assumption}[Finite carryover and non-anticipation]
\label{assu:carryover}
There is a carryover horizon $m>0$ such that, for any two assignment paths $\mathbf w$ and $\mathbf w'$,
\[
\mathbf w(s)=\mathbf w'(s)\text{ for every }s\in[t-m,t]
\quad\to\quad
Y_t(\mathbf w)=Y_t(\mathbf w').
\]
\end{assumption}

Let \(\mathbf W\) denote the random assignment path generated by \(D_G\), and let
\[
    Y_t:=Y_t(\mathbf W),
    \qquad
    E^a(t):=
    \mathbf 1\{\mathbf W(s)=a
    \text{ for all }s\in[t-m,t]\},
\]
be the observed outcome and pure exposure indicators.
To construct the estimator, we first need the probability of pure exposure under the design.
By stationarity and the renewal-reward theorem~\citep{ross1996stochastic},
the pure-exposure probability is
\begin{equation}
    p_G
    :=
    \PP[G]{E^a(t)=1}
    =
    \frac{\EE[G]{(R-m)_+}}{2\EE[G]{R}},
    \qquad a\in\{0,1\},
\label{eq:p_G}
\end{equation}
which is in general different from the marginal treatment assignment probability $1/2$ since pure exposure requires the current treatment to have been in place throughout the carryover period.
For any $G\in\mathcal G$ with $p_G>0$, we then consider the Horvitz-Thompson estimator\footnote{We view the stationary assignment process as extending before time \(0\), so that exposure is well-defined near the beginning of the observation window. Operationally, this amounts to starting the assignment process before outcome collection; alternatively, one may discard the initial carryover window.} \citep{horvitz1952generalization}
\begin{equation}
    \widehat\tau_T
    :=
    \frac{1}{T}\int_0^T
    \left\{
        \frac{E^1(t)Y_t}{p_G}
        -
        \frac{E^0(t)Y_t}{p_G}
    \right\}dt.
\label{eq:HT}
\end{equation}
This estimator is the continuous-time analogue of the inverse-probability-weighted estimators commonly used in design-based analyses of temporal experiments \citep{bojinov2019time,basse2023minimax,xiong2023bias,bojinov2020design}.

To formulate a robust design problem without imposing a parametric model on
the outcome paths, we restrict the adversarial outcome schedules to a bounded
class, as is common in minimax analyses of temporal experiments
\citep{bojinov2020design,ni2022balanced}.

\begin{assumption}[Bounded outcomes]
\label{assu:outcome}
There exists $B<\infty$ such that
\[
    |Y_t(\mathbf 1)|\le B,
    \qquad
    |Y_t(\mathbf 0)|\le B
\]
for almost every $t\in[0,T]$.
\end{assumption}

Let $\mathcal Y_B$ denote the class of potential-outcome schedules satisfying
Assumption~\ref{assu:outcome}. We then choose the duration distribution to minimize the worst-case mean squared error:
\[
    \inf_{\substack{G\in\mathcal G\\ p_G>0}}
    \sup_{Y\in\mathcal Y_B}
    \EE[G]{\bigl(\widehat\tau_T-\tau_T\bigr)^2}.
\]

\subsection{Error decomposition}

Having formulated the minimax design problem, the next question is how the
choice of duration distribution $G$ affects the estimation error. For the Horvitz-Thompson estimator,
we show that the error can be separated cleanly into two sources: how well pure treatment and pure control are balanced over time, and
how much pure exposure is available in the first place. This decomposition will
allow us to study these two aspects of the design separately and, in the next
section, connect them to the temporal structure induced by the duration
distribution.

On the outcome side, define the common outcome level and the half treatment
contrast by
\begin{subequations}
\begin{equation}
    u(t)
    :=
    \frac{Y_t(\mathbf 1)+Y_t(\mathbf 0)}{2},
    \qquad
    v(t)
    :=
    \frac{Y_t(\mathbf 1)-Y_t(\mathbf 0)}{2}.
\end{equation}
Intuitively, $u(t)$ captures variation of the baseline, while $v(t)$
captures variation of the treatment effect.
Meanwhile, on the design side, define the two processes
\begin{equation}
    X(t)
    :=
    \frac{E^1(t)-E^0(t)}{p_G},
    \qquad
    H(t)
    :=
    \frac{E^1(t)+E^0(t)}{p_G}-2,
\end{equation}
\end{subequations}
where $X(t)$ records whether time \(t\) is under pure treatment or pure control exposure, and $H(t)$ records whether time \(t\) is contaminated by the previous treatment run.

Because the design is stationary, its second-order dependence depends only on the time lag and can therefore be summarized by the autocovariance functions \citep{priestley1981spectral,brockwell1991time}
\begin{equation}
    \gamma_X(h)
    :=
    \operatorname{Cov}_G\!\bigl(X(t),X(t+h)\bigr),
    \qquad
    \gamma_H(h)
    :=
    \operatorname{Cov}_G\!\bigl(H(t),H(t+h)\bigr).
\end{equation}
Below, we show that these two covariance functions capture the two different ways
in which the assignment path contributes to estimation error. 

\begin{proposition}
\label{prop:decomp}
Under Assumptions~\ref{assu:carryover} and~\ref{assu:outcome}, for every $G\in\mathcal G$ with $p_G>0$,
\begin{subequations}
\begin{equation}
\label{eq:error_decomp}
    \widehat\tau_T-\tau_T
    =
    \frac{1}{T}\int_0^T
    \bigl\{u(t)X(t)+v(t)H(t)\bigr\}\,dt.
\end{equation}
Moreover,
\begin{equation}
\label{eq:errorsqd_decomp}
    \EE[G]{\p{\widehat\tau_T-\tau_T}^2}
    =
    \frac{1}{T^2}
    \int_0^T\int_0^T
    \left\{
        u(t)u(s)\gamma_X(t-s)
        +
        v(t)v(s)\gamma_H(t-s)
    \right\}
    \,ds\,dt.
\end{equation}
\end{subequations}
\end{proposition}

Proposition~\ref{prop:decomp} tells us where the estimation error comes from. On the one hand, the first term is driven by temporal imbalance between pure treatment and pure control exposure. When one treatment is overrepresented for a period of time, variation in the common outcome level \(u(t)\) can be mistaken for a treatment effect. 
On the other hand, the second term is driven by contamination. During contaminated periods, the treatment contrast cannot be learned directly, so the estimator must rely on pure-exposure periods to represent those missing periods.

This decomposition also shows why the duration distribution matters beyond its mean. This is because the variance is driven not only by how often switches happen on average, but also by how switch times are arranged over time. For example, nearly deterministic durations can create highly regular treatment–control imbalance, while a more dispersed duration distribution can produce occasional long stretches under the same treatment and irregular gaps between pure-exposure periods. These designs may have the same mean duration but very different temporal dependence in both $X$ and $H$. The full distribution of durations is therefore the natural object to optimize.

\subsection{A spectral representation}

The decomposition in Proposition~\ref{prop:decomp} shows that the design affects the estimation error
through the temporal dependence of the two processes $X$ and $H$. Since the design-induced processes $X$ and $H$ are stationary, a convenient way to describe this dependence is by asking
how strongly they fluctuate at different temporal frequencies. 
As we will see below, the resulting
spectral representation also leads to a simple tractable upper bound on the minimax
estimation error.

We restrict attention to designs for which the covariance functions of \(X\) and \(H\) are absolutely integrable. This excludes, in particular, designs with persistent periodic dependence such as alternating fixed blocks whose spectra contain point masses. For such designs, an adversary can choose a periodic outcome pattern aligned with a frequency at which the design has a spectral atom, so the resulting estimation error may not vanish as the experiment grows. 

\begin{assumption}[Absolute covariance integrability]
\label{assu:spectral}
The covariance kernels $\gamma_X$ and $\gamma_H$ are absolutely integrable on $\RR$.
\end{assumption}

Before introducing the spectral representation, it is useful to also make a simple change of time units. We measure time relative to the carryover horizon $m$ so that a carryover window of length \(m\) becomes a window of length one.
Correspondingly, a treatment run of duration \(R\) has normalized duration \(R/m\), and the experiment horizon becomes \(T/m\).
This change of units simply relabels the clock and does not alter the assignment or exposure structure at all.
However, it allows us to separate the physical time scale of the experiment from the shape of the design. 
After normalization, two designs with the same distribution of $R/m$ now have the same assignment structure on the normalized time scale regardless of the value of $m$. 
We can therefore work without loss of generality in normalized time with \(m=1\), and restore the original time scale afterward.

The same change of units carries over naturally to the frequency domain. 
Under this rescaling, a frequency $\omega$ on the original time scale corresponds to the normalized frequency $x := m\omega$ relative to the carryover horizon. We can then describe the temporal dependence of the design on this common scale through the normalized spectral densities
\[
    f_X(x)
    :=
    \int_{\mathbb R} \gamma_X(ms)e^{-ixs}\,ds,
    \qquad
    f_H(x)
    :=
    \int_{\mathbb R} \gamma_H(ms)e^{-ixs}\,ds.
\]
Similarly, define the normalized Fourier transforms of the outcome components by
\[
    U_T(x)
    :=
    \frac{1}{m}\int_0^T u(t)e^{-ixt/m}\,dt,
    \qquad
    V_T(x)
    :=
    \frac{1}{m}\int_0^T v(t)e^{-ixt/m}\,dt.
\]
We can now express both the exact estimation error and a tractable minimax bound entirely in terms of these normalized quantities.

\begin{theorem}
\label{theo:representation}
Under Assumptions~\ref{assu:carryover}--\ref{assu:spectral}, for any $G\in\mathcal G$ with $p_G>0$,
\begin{subequations}
\begin{equation}
\label{eq:representation}
\EE[G]{(\widehat\tau_T-\tau_T)^2}
    =
    \frac{m^2}{2\pi T^2}
    \int_{\mathbb R}
    \left\{
        f_X(x)|U_T(x)|^2
        +
        f_H(x)|V_T(x)|^2
    \right\}\,dx,
\end{equation}
and
\begin{equation}
\label{eq:proxy}
    \sup_{Y\in\mathcal Y_B}
    \EE[G]{(\widehat\tau_T-\tau_T)^2}
    \le
    \frac{B^2m}{T}\sup_{x\in\mathbb R}
    \max\{f_X(x),f_H(x)\}.
\end{equation}
\end{subequations}
Moreover, if $\sup_{x\in\mathbb R}\max\{f_X(x),f_H(x)\}=\max\{f_X(0),f_H(0)\}$,
then the bound is asymptotically sharp in the sense that
\begin{equation}
\label{eq:asyp_sharp}
    \lim_{T/m\to\infty}
    \sup_{Y\in\mathcal Y_B}
    \frac{T}{B^2m}
    \EE[G]{(\widehat\tau_T-\tau_T)^2}
    =
    \sup_{x\in\mathbb R}
    \max\{f_X(x),f_H(x)\}.
\end{equation}
\end{theorem}

\begin{figure}[t]
\centering
\includegraphics[width=0.91\textwidth]{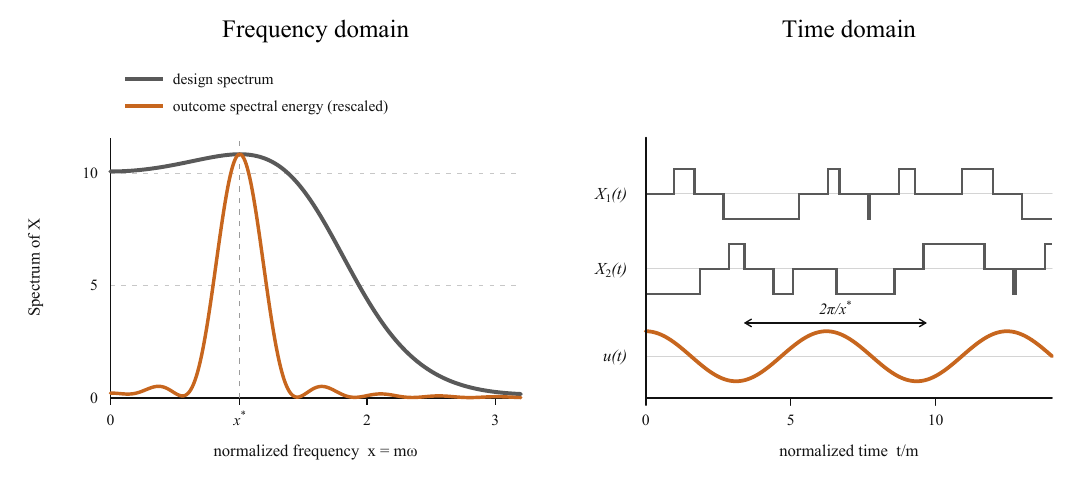}
\caption{Spectral vulnerability in the frequency and time domains. Left: the spectrum of the treatment--control imbalance process \(X\), together with the spectral energy of a periodic baseline-outcome pattern concentrated at the peak frequency \(x^\star=m\omega^\star\); the outcome spectral energy is rescaled for visualization. Right: two realizations of \(X(t)\) and the corresponding periodic outcome pattern. The peak frequency \(x^\star\) corresponds to a normalized time period of length \(2\pi/x^\star\), a time scale on which similar imbalance patterns tend to recur.}
\label{fig:spectral_overlap}
\end{figure}

Theorem~\ref{theo:representation} makes explicit the temporal patterns to which a given design is most vulnerable. 
Equation~\eqref{eq:representation} suggests that estimation error is large when the outcome energy spectrum \(\lvert U_T(x)\rvert^2\) overlaps with frequencies where \(f_X(x)\) is large, or when \(\lvert V_T(x)\rvert^2\) overlaps with frequencies where \(f_H(x)\) is large.
More concretely, a peak in \(f_X\) identifies a time scale on which treatment--control imbalance tends to recur. If the common outcome level \(u(t)\) varies on that same time scale, the imbalance can line up systematically with outcome variation. As a result, temporal variation in the common outcome level can be systematically mistaken for a treatment effect, leading to large estimation error. Likewise, a peak in \(f_H\) identifies a time scale on which the pattern of contaminated and uncontaminated observations tends to recur. If the treatment contrast \(v(t)\) varies on that same time scale, periods with systematically larger or smaller treatment effects can coincide with contamination, forcing the estimator to rely on other periods to represent them and again leading to large estimation error.

Figure~\ref{fig:spectral_overlap} makes this idea concrete for the \(X\)-component. 
Suppose the treatment--control imbalance spectrum \(f_X\) has a peak at \(x^\star\). A worst-case outcome schedule would then vary the common outcome level periodically at roughly the same frequency, or in other words place much of its energy spectrum \(|U_T(x)|^2\) near that peak. The left panel shows this overlap in the frequency domain.
The right panel shows what this frequency means in time: it corresponds to a period of \(2\pi/x^\star\) in units of the carryover horizon.
Indeed, the realized paths show a noisy tendency for similar imbalance patterns to reappear at roughly this time scale.
Although the realized \(X(t)\) paths are random, their dependence is particularly strong on this time scale. An outcome varying on a similar time scale can therefore line up with the design's treatment--control imbalance and produce large estimation error.

The exact minimax problem is nevertheless more difficult than simply locating the largest spectral peak. The outcome spectra \(U_T\) and \(V_T\) cannot be chosen independently frequency by frequency, because \(u\) and \(v\) are linked through the same pair of potential-outcome paths. The upper bound~\eqref{eq:proxy} in Theorem~\ref{theo:representation} bypasses this difficulty by guarding against the largest value across frequencies of either design spectrum. 
When the largest spectral value occurs at zero frequency (i.e., nonoscillating patterns), Theorem~\ref{theo:representation} shows that the bound becomes asymptotically exact. 
More generally, the criterion is conservative because it effectively allows the outcome to place all of its admissible variation at the most unfavorable frequency, which need not be attainable by a bounded time-domain outcome path. Nevertheless, it provides a tractable way to compare and optimize duration distributions.\footnote{It is also possible to obtain a worst-case upper bound $B^2/T\cdot \max\left\{\int \abs{\gamma_X(h)}\,dh,\int \abs{\gamma_H(h)}\,dh\right\}$ directly from Proposition~\ref{prop:decomp}. However, this bound can be substantially looser when the autocovariances have mixed signs across lags, because it ignores cancellations between positive and negative autocovariances at different lags.} 
As we show in the next section, even a simple two-rate random-duration design yields a substantially tighter worst-case mean-squared-error certificate than the regular-switchback benchmark.

\begin{remark}
The zero-frequency condition in Theorem~\ref{theo:representation} is a consequence of the pointwise boundedness constraint imposed in Assumption~\ref{assu:outcome}. If the adversary were instead allowed to choose any outcome schedule satisfying the second-moment condition
\[
    \frac1T\int_0^T Y_t(\mathbf a)^2\,dt\le B^2, \qquad a=0,1,
\]
then the spectral upper bound~\eqref{eq:proxy} would always be asymptotically sharp. 
In that case, an appropriately scaled sinusoidal outcome can concentrate its asymptotic spectral energy at the maximizing frequency while exhausting the second-moment constraint. 
Consequently, minimizing the proxy in~\eqref{eq:proxy} solves the exact asymptotic minimax design problem under this outcome class, while the proxy remains a valid conservative certificate under the stronger pointwise boundedness assumption.
We give the formal characterization in Appendix~\ref{sec:l2_sharpness}.
\end{remark}

\section{Optimizing the random-duration design}

Theorem~\ref{theo:representation} suggests a tractable way to compare random-duration
designs. Define
\begin{equation}
    J(G)
    :=
    \sup_{x\in\mathbb R}
    \max\{f_X(x),f_H(x)\}.
\end{equation}
The quantity $J(G)$ controls the worst-case estimation error uniformly over the
bounded outcome class, and is asymptotically exact whenever the largest spectral
value occurs at zero frequency. We therefore use $J(G)$ as our design criterion
and ask how the duration distribution $G$ should be chosen to make it small.

Optimizing over all possible duration distributions is an infinite-dimensional problem.
To simplify the problem, we proceed in two steps. First, we exploit the renewal structure in Definition~\ref{defi:renewal} to show that the spectra \(f_X\) and \(f_H\), and thus \(J(G)\), can be computed directly from the duration distribution \(G\). We then restrict attention to a simple three-parameter family in which the switching hazard takes one constant rate early in a run and another after a single changepoint. This is perhaps the simplest parametric family that allows the switching behavior to depend on run age while remaining easy to implement and optimize. As we will see, optimizing even this simple family produces a markedly different duration distribution from conventional fixed-block designs and a substantially tighter worst-case estimation-error guarantee.

\subsection{Computing the spectral criterion}

Although \(J(G)\) is defined through the spectra of the full processes \(X\) and \(H\), the renewal structure makes these spectra much easier to compute. 
Let $Z:=R/m$ denote the normalized duration. In a random-duration design, the entire assignment path is generated by i.i.d. durations \(Z_1,Z_2,\ldots\) together with alternating treatment labels. Therefore, to understand the temporal dependence of \(X\) and \(H\), we only need to understand how the process behaves within a realized run and how different runs are positioned relative to one another over time.

At frequency \(x\), define
\begin{equation}
    \phi_G(x):=\EE[G]{e^{ixZ}},
    \qquad
    \psi_Z(x):=
    \mathbf 1\{Z>1\}\int_1^Z e^{-ixs}\,ds.
\end{equation}
Here, \(\psi_Z(x)\) describes the within-run pattern at frequency \(x\) for a realized run of duration \(Z\). It is the Fourier transform of the interval from run age \(1\) to \(Z\) (i.e., the uncontaminated portion of the run) when that interval exists. In contrast, \(\phi_G(x)\) summarizes the distribution of the random durations at frequency \(x\). Since the distance between different runs is formed by sums of independent durations, the dependence across runs is governed by powers of \(\phi_G(x)\). Below, we show that the spectra \(f_X\) and \(f_H\) can be obtained directly using these within-run and across-run quantities.

\begin{proposition}
\label{prop:renewal_spectral}
Under Assumption~\ref{assu:spectral}, assume in addition that $G$ has a density, $\EE[G]{Z^2}<\infty$, and $p_G>0$.
Then, the normalized spectral densities of $X$ and $H$ can be written as
\[
f_X(x)
=
\begin{cases}
\displaystyle
\frac{1}{p_G^2\EE[G]{Z}}
\left[
    \EE[G]{|\psi_Z(x)|^2}
    -
    2\recal\cb{
        \frac{
            \EE[G]{\psi_Z(x)e^{ixZ}}
            \EE[G]{\overline{\psi_Z(x)}}
        }{
            1+\phi_G(x)
        }
    }
\right],
& x\neq 0,\\[16pt]
\displaystyle
\frac{
    \Var[G]{(Z-1)_+}
}{
    p_G^2\EE[G]{Z}
},
& x=0,
\end{cases}
\]
and
\[
f_H(x)
=
\begin{cases}
\displaystyle
\frac{1}{p_G^2\EE[G]{Z}}
\left[
    \EE[G]{|\psi_Z(x)|^2}
    +
    2\recal\left\{
        \frac{
            \EE[G]{\psi_Z(x)e^{ixZ}}
            \EE[G]{\overline{\psi_Z(x)}}
        }{
            1-\phi_G(x)
        }
    \right\}
\right],
& x\neq 0,\\[16pt]
\displaystyle
\frac{
    \Var[G]{(Z-1)_+-2p_G Z}
}{
    p_G^2\EE[G]{Z}
},
& x=0.
\end{cases}
\]
\end{proposition}

Proposition~\ref{prop:renewal_spectral} separates two aspects of the temporal dependence induced by a random-duration design. In each spectrum, the first term reflects variation within individual renewal intervals, while the second captures dependence across different intervals. The latter is governed by the duration distribution through its characteristic function \(\phi_G\), which summarizes how the random spacing between successive runs behaves at each frequency. The difference between \(f_X\) and \(f_H\) comes mainly from the alternating treatment labels, as the same sequence of renewal intervals enters the two processes with different sign patterns. In Appendix~\ref{sec:num_opt}, we show that the terms involving $\psi_Z$ can be further expressed entirely in terms of the truncated characteristic function and the tail probability.

With Proposition~\ref{prop:renewal_spectral}, we are now able to compute \(J(G)\) directly from the duration distribution and optimize it over a parametric family of random-duration designs.
With this characterization, evaluating a random-duration design no longer requires analyzing or simulating the full assignment path. 
Once the duration distribution \(G\) is specified, both spectra are determined by one-dimensional expectations involving the normalized duration \(Z\). These expectations can be evaluated analytically when the duration law is simple, or by numerical integration otherwise. 
We next apply this characterization to a simple two-rate family in which the switching hazard changes once with the age of the current run.

\subsection{A two-rate random-duration design}
\label{sec:two_rate_defi}

Although the duration distribution $G$ is the underlying design object, it is
often more intuitive to describe a duration distribution with density through
its switching hazard. For a duration
$R\sim G$, define
\begin{equation*}
    \lambda_G(r)
    :=
    \lim_{\Delta\downarrow0}
    \frac{1}{\Delta}
    \PP[G]{R\in[r,r+\Delta)\mid R\ge r},
    \qquad r\ge0.
\end{equation*}
Thus, conditional on the current treatment run having reached age $r$,
$\lambda_G(r)\Delta$ is approximately the probability that it ends during the
next $\Delta$ units of time.
This is the continuous-time analogue of the conditional
switching probability $q_G$ we discussed in Section~\ref{sec:special_case}.

We focus on a particularly simple family in which the switching rate can change with run
age: two constant rates separated by a single changepoint,
\begin{equation}
\label{eq:two_rate_hazard}
    \lambda_{c_1,c_2,\theta}(r)
    =
    \begin{cases}
        c_1/m, & 0\le r\le \theta m,\\[3pt]
        c_2/m, & r>\theta m,
    \end{cases}
    \qquad c_1\ge 0,\,c_2,\theta>0,
\end{equation}
where $c_1$ controls the switching rate early in a run, $c_2$ controls the switching rate later in
the run, and $\theta m$ determines when the switching rule changes. 

The corresponding survival function is then
\[
    \PP[G]{R>r}
    =
    \begin{cases}
        \exp(-c_1 r/m), & 0\le r\le \theta m,\\[3pt]
        \exp\!\left\{-c_1\theta-c_2(r/m-\theta)\right\},
            & r>\theta m.
    \end{cases}
\]
The duration therefore has a piecewise-exponential survival law, with decay rate \(c_1/m\) before the changepoint and \(c_2/m\) afterward. Thus, the three parameters $(c_1,c_2,\theta)$ completely characterize the duration distribution, and,
through its induced spectra, the design criterion $J(G)$. Since these duration distributions are nonlattice with exponentially decaying tails, the induced renewal processes satisfy Assumption~\ref{assu:spectral}.

We now minimize the spectral criterion over the three-parameter family in
\eqref{eq:two_rate_hazard} using the characterization in Proposition~\ref{prop:renewal_spectral}. Let $G_{c_1,c_2,\theta}$ denote the duration distribution induced by
\eqref{eq:two_rate_hazard}. We solve
\[
    \inf_{c_1\ge 0,c_2,\theta>0}
    J\!\left(G_{c_1,c_2,\theta}\right).
\]
By Proposition~\ref{prop:renewal_spectral}, for each $(c_1,c_2,\theta)$, the
two spectra can be evaluated directly from one-dimensional expectations under
$G_{c_1,c_2,\theta}$. Solving this problem numerically yields
\begin{equation}
\label{eq:two_rate_solution}
    c_1^\star \approx 0.182,
    \qquad
    c_2^\star \approx 0.885,
    \qquad
    \theta^\star \approx 1.045,
\end{equation}
and the corresponding spectral criterion is $J(G^\star) \approx 10.843$. More details about the numerical minimization can be found in Appendix~\ref{sec:num_opt}.

The optimized switching rule has a clear structure. The switching rate is much
lower early in a run than later, so the design tends to keep a treatment in
place early and becomes substantially more willing to switch as the run gets
older. 
Despite optimizing a conservative spectral proxy rather than the exact minimax error, the optimal changepoint lies strikingly close to \(\theta=1\), the boundary between the contaminated and uncontaminated portions of a run.
At the same time, the optimal early switching rate is small but strictly positive. Thus, it is not optimal, at least within this two-rate family, to protect every run until it reaches pure exposure. The design deliberately sacrifices some runs before age \(m\), which suggests that a hard minimum duration may introduce too much regularity. The optimum instead balances preserving the chance of reaching pure exposure against retaining randomness in the switching times.
Algorithm~\ref{alg:two_rate} summarizes how to run the optimized two-rate
design and estimate the global average treatment effect ~\eqref{eq:GATE} in practice. 

\begin{algorithm}[t]
\caption{Two-rate random-duration switchback}
\label{alg:two_rate}
\begin{algorithmic}[1]
\Statex \textbf{Input:} Carryover horizon $m>0$ and experimentation horizon $T>0$.
\Statex \textbf{Parameters:} $(c_1,c_2,\theta)=(0.182,0.885,1.045)$.
\Statex \textbf{Duration rule:} Draw $R_1\sim\operatorname{Exp}(c_1/m)$.
If $R_1\le\theta m$, use $r=R_1$; otherwise draw
$R_2\sim\operatorname{Exp}(c_2/m)$ and use $r=\theta m+R_2$.
\State At time $-m$, assign treatment or control with equal probability.
\State Draw $r$ by the duration rule. Accept it if $r\le\theta m$;
otherwise accept with probability $c_1/c_2$.{\footnotemark }
Repeat until a draw is accepted and schedule the first switch at $-m+r$.
\State At each scheduled switch time $t<T$, switch to the other arm,
draw a fresh duration $r$ by the duration rule,
and schedule the next switch at $t+r$. End the experiment at $T$.
\State Compute the Horvitz-Thompson estimator based on~\eqref{eq:HT}, with the pure-exposure probability $p_G$ evaluated from~\eqref{eq:p_G} (see Table~\ref{tab:design_summary} for an approximate value).
\end{algorithmic}
\end{algorithm}

\footnotetext{
To initialize the stationary renewal process, the time to the first switch
needs to have density proportional to $\PP[G]{R>r}$
\citep[Section~3.5]{ross1996stochastic}. For the two-rate duration law, accepting draws after $\theta m$ with probability $c_1/c_2$ gives exactly a density proportional to $\PP[G]{R>r}$.
This is equivalent to the length-biased initialization described in
Section~\ref{sec:renewal}.}

The proximity of $\theta^\star$ to one also suggests a simpler implementation.
If we fix the changepoint exactly at the carryover horizon, $\theta=1$, and
optimize only over the two switching rates, we obtain
\[
    c_1^\star \approx 0.138,
    \qquad
    c_2^\star \approx 0.861,
    \qquad
    J(G^\star_{\theta=1})\approx10.854.
\]
Comparing with the performance of the free-change-point design~\eqref{eq:two_rate_solution}, 
the spectral criterion of this design increases by only about $0.1\%$. Thus, 
if a practitioner finds it easier to implement the simpler two-parameter specification that changes the switching rate exactly at the carryover horizon, this restriction incurs essentially no loss in the worst-case certificate.

\subsection{Comparison with fixed-block designs}
\label{sec:comparison}

We now compare the optimized random-duration design with the fixed-block designs from
Section~\ref{sec:special_case}. 
From a spectral viewpoint, the alternating fixed-block design has an immediate vulnerability: the design has spectral atoms at its periodic frequencies. Once the initial phase and treatment label are chosen, the assignment pattern is perfectly periodic. A bounded outcome pattern with the same period can therefore align perfectly with the treatment--control imbalance, so the worst-case estimation error need not decay as the time horizon $T$ grows. As a result, whenever $\ell>m$ so that the Horvitz-Thompson estimator is well-defined, there exists a bounded periodic outcome schedule for which the estimation error does not vanish:
\begin{equation*}
\liminf_{T\to\infty} 
    \EE[G]{(\widehat\tau_T-\tau_T)^2} > 0.
\end{equation*}
Thus, $\widehat\tau_T$ may not even be consistent when using the alternating fixed-block design.

The regular switchback removes this exact periodicity by randomizing each block independently. It therefore provides the more meaningful quantitative benchmark for our random-duration designs. 
To choose its block length, we follow the minimax design of~\cite{bojinov2020design}, which studies the analogous discrete-time problem under finite carryover using the corresponding Horvitz-Thompson estimator. Their optimal design uses blocks of length \(m\) away from the experiment boundaries, which motivates the choice \(\ell=m\) in our stationary formulation.

Unlike the
continuously distributed duration laws considered in Proposition~\ref{prop:renewal_spectral}, the
regular switchback has a discrete duration distribution, but its spectra can
be calculated directly. 
We give its spectral criterion in the following corollary.

\begin{corollary}
\label{coro:regular}
Under the regular switchback with block length $\ell=m$,
\begin{subequations}
\begin{equation}
    \sup_{Y\in\mathcal Y_B}
    \EE[G_{\mathrm{reg}}]{(\widehat\tau_T-\tau_T)^2}
    \le
    \frac{16B^2m}{T},
\end{equation}
and this bound is asymptotically sharp:
\begin{equation}
    \lim_{T/m\to\infty}
    \frac{T}{B^2m}
    \sup_{Y\in\mathcal Y_B}
    \EE[G_{\mathrm{reg}}]{(\widehat\tau_T-\tau_T)^2}
    =
    16.
\end{equation}
\end{subequations}
\end{corollary}

Corollary~\ref{coro:regular} gives an asymptotically sharp benchmark for the gain from
optimizing the duration distribution. The regular switchback has asymptotic
worst-case constant $16$, whereas our optimized random-duration design has the
conservative upper bound $J(G^\star)\approx 10.843$. Thus, the optimized two-rate
design reduces the asymptotic worst-case mean squared error by at least $32.2\%$
relative to the regular switchback.

For the regular switchback, the binding component is $f_X(0)$. Since the
spectral maximum occurs at zero frequency, the upper bound in
Theorem~\ref{theo:representation} is asymptotically sharp. This also aligns with the
least-favorable outcome construction of \citet{bojinov2020design}, where the
outcomes under global treatment and global control are equal and constant over
time, at either $B$ or $-B$. In our decomposition, this corresponds to
$v(t)=0$ and $u(t)=\pm B$, which places all outcome variation at zero frequency in
the treatment--control imbalance channel.

\section{Numerical illustration}
\label{sec:sim}

We conclude with a simple numerical illustration of how the different duration distributions translate into finite-horizon estimation performance. Throughout, we normalize the carryover horizon to \(m=1\) and compare four designs: alternating fixed blocks, the regular switchback, an optimized one-rate design, and the optimized two-rate design from Section~\ref{sec:two_rate_defi}.
We use the same Horvitz-Thompson estimator for all designs, and initialize each assignment process as described in Section~\ref{sec:renewal}.

For alternating fixed blocks, we take \(\ell=2m\); with this choice, it has the same mean duration and pure-exposure probability as the regular switchback benchmark. 
For the regular switchback design, we take \(\ell=m\) following the discussion in Section~\ref{sec:comparison}. 
We also consider a one-rate random-duration design with constant switching hazard \(\lambda(r)=c/m\), which removes the fixed switching lattice but does not allow switching behavior to depend on run age. 
This one-rate design has exponential treatment durations and is the continuous-time analogue of the symmetric memoryless Markov switchbacks considered by \citet[Definition 15.2]{wager2026causal} and \citet{guo2026experimental}.
We then choose \(c\) to minimize \(J(G)\), which yields the optimized rate $c^*\approx 0.768$. 
Finally, we consider the optimized two-rate design from Section~\ref{sec:two_rate_defi}, which allows the switching rate to change with run age.

\subsection{Duration distributions}
\label{sec:sim_design}

\begin{figure}[t]
\centering
\includegraphics[width=0.49\textwidth]{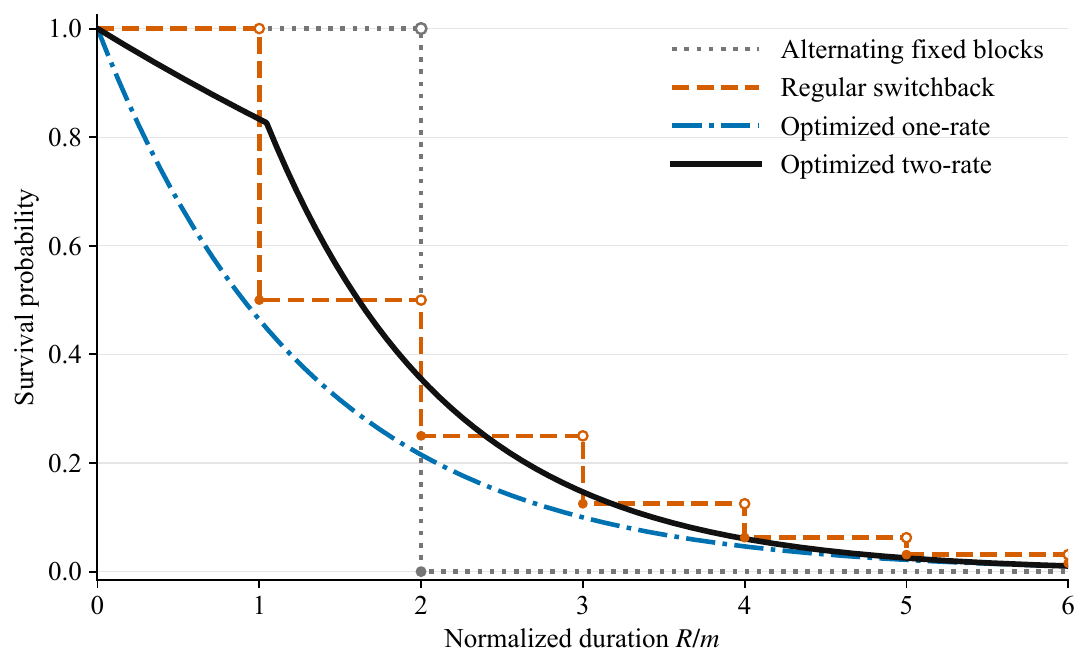}
\includegraphics[width=0.49\textwidth]{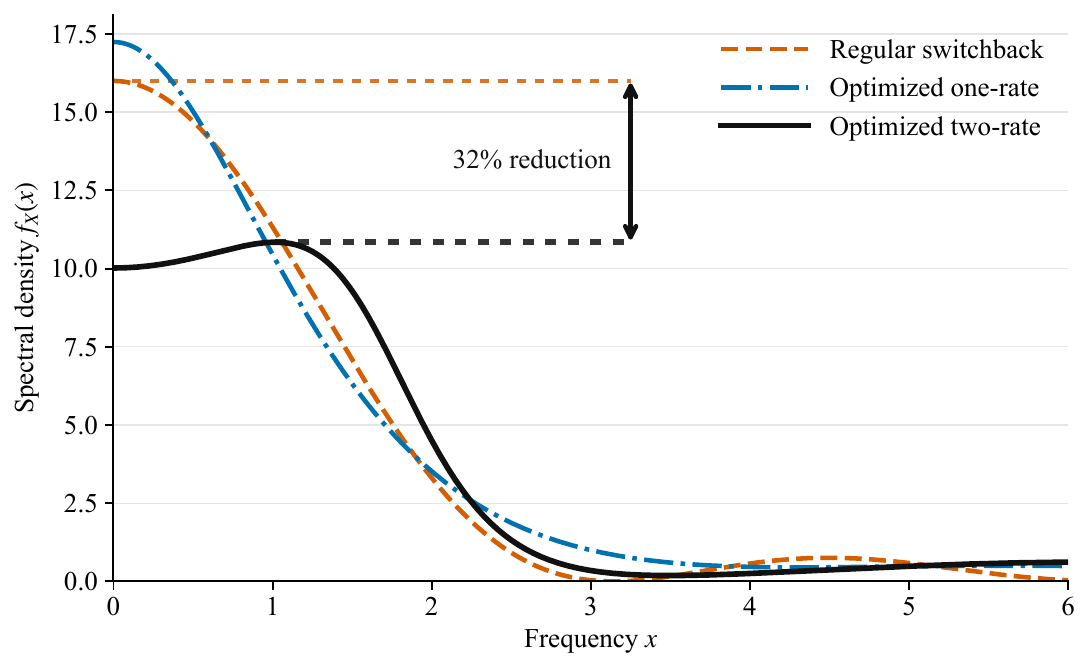}
\caption{Duration distributions and treatment–control imbalance spectra. Left: survival functions of the normalized duration \(R/m\) under the four designs. Right: normalized spectral density \(f_X(x)\) for the three randomized designs; alternating fixed blocks are omitted because their spectrum contains point masses rather than an ordinary spectral density.}
\label{fig:run_length_comparison}
\end{figure}

We first compare how the four design rules translate into different duration distributions. The left panel of Figure~\ref{fig:run_length_comparison} shows the survival function of the normalized duration \(R/m\). Alternating fixed blocks place all mass at \(R/m=2\), while the regular switchback produces a geometric distribution supported on the lattice \(1,2,\ldots\). The one-rate design removes this lattice structure but retains a memoryless exponential shape. In contrast, the optimized two-rate design suppresses switching early in a run and accelerates it later. As a result, its survival probability remains high through the carryover horizon and then declines more rapidly afterward.

These differences are not well summarized by the average duration or by the amount of pure exposure alone. As shown in Table~\ref{tab:design_summary}, the regular switchback and optimized two-rate design have very similar mean durations, \(2.00m\) and \(1.89m\), and nearly identical pure-exposure probabilities, \(0.250\) and \(0.258\). 
Yet these similar coarse summaries do not translate into similar spectral performance.
The regular switchback has a worst spectral vulnerability \(J(G)=16\), while the optimized two-rate design reduces it to \(10.84\). The one-rate benchmark helps clarify where this gain comes from. Although it removes the fixed switching lattice, its spectral criterion is actually even larger than that of the regular switchback. The improvement therefore comes not simply from randomizing switch times continuously, but from allowing the switching rule to depend on the age of the current run.

\begin{table}[t]
\centering
\begin{tabular}{lccc}
\toprule
Design & $\EE[G]{R/m}$ & $p_G$ & $J(G)$ \\
\midrule
Alternating fixed blocks      & 2.000 & 0.250 & --- \\
Regular switchback            & 2.000 & 0.250 & 16.000 \\
Optimized one-rate    & 1.302 & 0.232 & 17.244 \\
Optimized two-rate    & 1.886 & 0.258 & 10.843 \\
\bottomrule
\end{tabular}
\caption{Mean normalized duration, pure-exposure probability, and spectral criterion for the four designs.}
\label{tab:design_summary}
\end{table}

The right panel of Figure~\ref{fig:run_length_comparison} compares the treatment--control imbalance spectrum \(f_X\), which is the binding component of \(J(G)\) for all three designs shown. 
The regular switchback and the one-rate design both have pronounced vulnerability at low frequencies. This similarity is consistent with their common memoryless structure: the former has geometric durations and the latter exponential durations, so the chance of switching does not increase with the age of the current run. As a result, both designs have highly variable run durations, which can contribute substantially to spectral mass at low frequencies.
The optimized two-rate design substantially reduces this zero-frequency vulnerability by suppressing early switches and accelerating later ones, with its spectral mass spread more evenly across frequencies.

\subsection{Finite-horizon estimation error}

We next examine how the differences in temporal dependence above translate into
finite-horizon estimation error under a simple causal model with carryover. In temporal
experiments, recurrent changes in the underlying system can affect both the baseline
outcome and the effect of treatment. For example, demand or congestion may follow daily
or weekly cycles, while the value of an intervention may also vary over the same cycle.
We therefore let the common outcome level and treatment contrast vary with the same
recurrent pattern, with the treatment contrast varying more mildly.

Recurrent outcome patterns can also be more complex than a single sinusoid, with
multiple peaks or fluctuations within the same cycle. To capture this while keeping the
outcome model transparent, we consider a periodic baseline with fundamental frequency
$\nu$ and its first two harmonics:
\[
b_\nu(t)
=
\frac{1}{3}
\left\{
\sin\left(\nu(t-T/2)\right)
+
\sin\left(2\nu(t-T/2)\right)
+
\sin\left(3\nu(t-T/2)\right)
\right\}.
\]
We vary $\nu$ over a dense grid on $(0,6]$, which covers recurrent patterns from
very slow variation near $\nu=0$ to periods as short as $2\pi/6$
on the normalized time scale.

We define the potential outcome under an assignment path $\mathbf w$ by
\[
Y_t(\mathbf w)
=
b_\nu(t)
+
\left\{\frac12+\frac14 b_\nu(t)\right\}
\left[
\mathbf w(t)-\frac12
+
\int_0^1
2(1-s)\left\{\mathbf w(t-s)-\frac12\right\}\,ds
\right].
\]
The treatment has both an immediate effect and a carryover effect that decays linearly
over the carryover horizon. Under global control and global treatment,
\[
Y_t(\mathbf 1)
=
b_\nu(t)+\frac12+\frac14b_\nu(t),
\qquad
Y_t(\mathbf 0)
=
b_\nu(t)-\frac12-\frac14b_\nu(t).
\]
The treatment effect is therefore
\[
Y_t(\mathbf 1)-Y_t(\mathbf 0)
=
1+\frac12 b_\nu(t),
\]
so it varies with the same recurrent pattern as the baseline, but with smaller amplitude.
Since $b_\nu$ has zero time average over $[0,T]$, $\tau_T=1$
for every value of $\nu$.

We set \(T=50\) and vary \(\nu\) over the grid \(\{0.01,0.02,\ldots,6\}\). For each value of \(\nu\) and each design, we treat the resulting potential-outcome schedule as fixed.
For each design $G$, we then repeatedly draw assignment paths and compute the
Horvitz-Thompson estimator. 
To numerically evaluate the continuous-time estimator, we discretize time using a regular grid with spacing \(0.01\), and the results are essentially unchanged when the spacing is halved to \(0.005\).
The mean squared error for this fixed schedule is $\EE[G]{(\widehat\tau_T-1)^2}$,
where the expectation is only over the treatment assignment. We estimate this quantity
by averaging the squared errors across 4,000 independent assignment realizations.

\begin{figure}[t]
\centering
\includegraphics[width=0.75\textwidth]{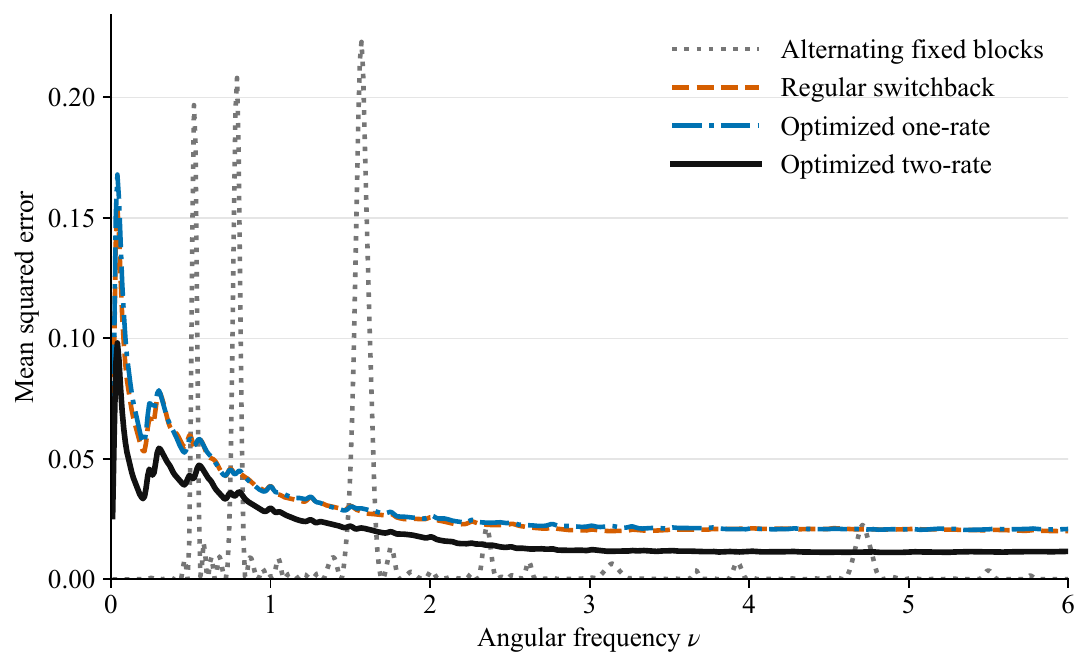}
\caption{Finite-horizon mean squared error as a function of the outcome frequency \(\nu\) under the four experimental designs, with carryover horizon \(m=1\) and experiment horizon \(T=50\).}
\label{fig:frequency_sweep}
\end{figure}

Figure~\ref{fig:frequency_sweep} shows that alternating fixed blocks behave very differently from the randomized designs. They achieve very small error for most frequencies, but exhibit large spikes around a discrete set of frequencies. These spikes arise when one of the harmonics in the outcome pattern aligns with a periodic frequency of the assignment process. Thus, deterministic alternation can perform well when the outcome frequencies do not align with the periodic switching pattern, but its performance is highly sensitive to recurrent outcome patterns that do.

The other three designs do not exhibit these large spikes. The regular switchback and optimized one-rate design behave very similarly as \(\nu\) varies. This is consistent with their common memoryless structure. Removing the fixed switching lattice therefore changes little by itself. In contrast, the optimized two-rate design has substantially lower mean squared error over the full range of frequencies considered, with the largest gains at lower frequencies where the other randomized designs are most vulnerable.
This suggests that run-age-dependent switching can substantially reduce estimation error across a broad range of temporal outcome patterns.


\bibliography{references}
\bibliographystyle{plainnat}

\newpage
\appendix
\begin{center}
\textbf{\Large Supplemental Materials} \\ 
\end{center}

\setcounter{equation}{0}
\setcounter{figure}{0}
\setcounter{table}{0}
\setcounter{page}{1}
\makeatletter
\renewcommand{\theequation}{S\arabic{equation}}
\renewcommand{\thefigure}{S\arabic{figure}}
\renewcommand{\thetable}{S\arabic{table}}
\renewcommand{\bibnumfmt}[1]{[S#1]}
\renewcommand{\citenumfont}[1]{S#1}

\section{Asymptotic sharpness under a second-moment constraint}
\label{sec:l2_sharpness}

In this section, we show that the spectral criterion has a stronger minimax interpretation under a natural second-moment restriction on the outcome schedules. 
In particular, if the adversary is constrained by its time-averaged second moment, we show that the spectral upper bound is asymptotically attained regardless of where its maximum occurs.

\begin{assumption}[Bounded second-moment averages]
\label{assu:l2_avg}
There exists $B<\infty$ such that
\begin{equation}
     \frac1T\int_0^T Y_t(\mathbf a)^2\,dt\le B^2, \qquad a=0,1.
\end{equation}
\end{assumption}

This condition can be viewed as the continuous-time analogue of a bounded finite-population second-moment condition, 
\[
    \frac{1}{T}\sum_{t=1}^T Y_t(\mathbf a)^2 \le B^2, \qquad a=0,1.
\]
Rather than requiring outcomes to be uniformly bounded at every time as in Assumption~\ref{assu:outcome}, it controls their average squared magnitude over the experimental horizon.

Let $\mathcal Y_{B,2}(T)$ denote the class of potential-outcome schedules satisfying
Assumption~\ref{assu:l2_avg}. Below, we show that, under this outcome class, the spectral criterion is no longer merely an upper bound. Instead, it gives the exact asymptotic worst-case constant.

\begin{proposition}
\label{prop:l2_sharpness}
Under Assumptions~\ref{assu:carryover},~\ref{assu:spectral} and~\ref{assu:l2_avg}, suppose in addition that $p_G>0$. Then,
\begin{equation}
\lim_{T/m\to\infty}
\sup_{Y\in\mathcal Y_{B,2}(T)}
\frac{T}{B^2m}
\EE[G]{(\widehat\tau_T-\tau_T)^2}
=
J(G).
\end{equation}
\end{proposition}

Proposition~\ref{prop:l2_sharpness} shows that the spectral criterion is not merely a convenient upper bound. Under the bounded second-moment-average condition, \(J(G)\) is exactly the asymptotic worst-case risk of design \(G\). Consequently, over any candidate class of designs, minimizing \(J(G)\) is equivalent to solving the corresponding asymptotic minimax design problem. In particular, the two-rate design obtained in Section~\ref{sec:two_rate_defi} by minimizing \(J(G)\) has an exact minimax interpretation under this outcome class, while the same criterion remains a valid conservative certificate under the stronger pointwise boundedness assumption.

\section{Numerical optimization of the two-rate design}
\label{sec:num_opt}

We carry out all numerical calculations on the normalized time scale $m=1$.
For parameters $(c_1,c_2,\theta)$, the normalized duration $Z=R/m$ has
density
\[
g_{c_1,c_2,\theta}(z)
=
\begin{cases}
c_1 e^{-c_1 z}, & 0\le z\le\theta,\\
c_2 e^{-c_1\theta-c_2(z-\theta)}, & z>\theta.
\end{cases}
\]
Its mean is
\[
\mu:=\EE[G]{Z}
=
\frac{1-e^{-c_1\theta}}{c_1}
+
\frac{e^{-c_1\theta}}{c_2},
\]
and
\[
p_G
=
\frac{\EE[G]{(Z-1)_+}}{2\mu}.
\]

For each parameter triple, we evaluate the two spectra in
Proposition~\ref{prop:renewal_spectral}. At nonzero frequency $x$, define the characteristic function and the truncated characteristic function
\[
\phi_G(x)=\EE[G]{e^{ixZ}},
\qquad
M_G(x)=\EE[G]{e^{ixZ}\1\p{Z>1}}.
\]
For the two-rate piecewise exponential distribution, both quantities are available in closed
form. In particular,
\[
\phi_G(x)
=
\frac{c_1\{1-e^{-(c_1-ix)\theta}\}}{c_1-ix}
+
\frac{c_2e^{-c_1\theta+ix\theta}}{c_2-ix}.
\]
The truncated characteristic function $M_G(x)$ is
\[
M_G(x)
=
\begin{cases}
\displaystyle
\frac{c_1\{e^{-(c_1-ix)}
-e^{-(c_1-ix)\theta}\}}{c_1-ix}
+
\frac{c_2e^{-c_1\theta+ix\theta}}{c_2-ix},
& \theta\ge1,\\[3mm]
\displaystyle
\frac{c_2
e^{-c_1\theta-c_2(1-\theta)+ix}}
{c_2-ix},
& \theta<1.
\end{cases}
\]
Recall
\[
\psi_Z(x)
=
\1\p{Z>1}
\frac{e^{-ix}-e^{-ixZ}}{ix}.
\]
The expectations appearing in Proposition~\ref{prop:renewal_spectral}
can be evaluated as
\[
\EE[G]{\psi_Z(x)e^{ixZ}}
=
\frac{e^{-ix}M_G(x)-\PP[G]{Z>1}}{ix},
\qquad
\EE[G]{\overline{\psi_Z(x)}}
=
\frac{M_G(x)-e^{ix}\PP[G]{Z>1}}{ix},
\]
\[
\EE[G]{|\psi_Z(x)|^2}
=
\frac{2}{x^2}
\left\{
\PP[G]{Z>1}-\recal\!\left(e^{-ix}M_G(x)\right)
\right\}.
\]
We substitute these expressions into Proposition~\ref{prop:renewal_spectral}
to obtain $f_X(x)$ and $f_H(x)$. At $x=0$, we use the corresponding
zero-frequency formulas in Proposition~\ref{prop:renewal_spectral}
directly.

Since both spectra are symmetric around zero, it suffices to maximize over $x\ge0$.
For each $(c_1,c_2,\theta)$, we first evaluate each spectrum on a dense
frequency grid over $[0,20]$, identify all grid-local maxima, and refine
each candidate using one-dimensional bounded optimization. We include
$x=0$ separately as a candidate. Increasing the frequency range to
$[0,100]$ does not change the reported maxima.

We then minimize
\[
J(G_{c_1,c_2,\theta})
=
\max\left\{
\sup_{x\ge0}f_X(x),
\sup_{x\ge0}f_H(x)
\right\}
\]
over the design parameters. We first optimize over
$(\log c_1,\log c_2,\log\theta)$ to enforce positivity, and then separately optimize over the boundary $c_1=0$. We use a global differential-evolution
search followed by a local Nelder--Mead refinement. Repeating the search
from multiple random seeds and over enlarged parameter ranges gives the
same solution to the reported precision.

For the unrestricted three-parameter family, this procedure gives
\[
c_1^\star=0.18237,\qquad
c_2^\star=0.88483,\qquad
\theta^\star=1.04496,
\]
with
\[
J(G^\star)=10.84281.
\]
The binding component is $f_X$, whose maximum occurs near
$x^\star=1.0161$. The corresponding design has $p_{G^\star}=0.25759$.

Fixing the changepoint at $\theta=1$ and repeating the two-parameter
optimization gives
\[
c_1^\star=0.13811,\qquad
c_2^\star=0.86128,
\qquad
J(G^\star_{\theta=1})=10.85438,
\]
where the binding $X$-spectrum reaches its maximum at
$x^\star=1.00884$, with $p_{G^\star_{\theta=1}}=0.25993$.

\section{Proof}

\subsection{Proof of Proposition~\ref{prop:decomp}}

We first prove the error identity~\eqref{eq:error_decomp}.
On the event $E^a(t)=1$, the assignment path agrees with the constant
path $\mathbf a$ throughout $[t-m,t]$. By Assumption~\ref{assu:carryover},
\[
E^a(t)Y_t
=
E^a(t)Y_t(\mathbf a).
\]
Therefore,
\begin{align*}
\widehat\tau_T-\tau_T
&=
\frac{1}{T}\int_0^T
\left\{
\frac{E^1(t)}{p_G}Y_t(\mathbf 1)
-
\frac{E^0(t)}{p_G}Y_t(\mathbf 0)
-
Y_t(\mathbf 1)+Y_t(\mathbf 0)
\right\}dt.
\end{align*}
Since
\[
Y_t(\mathbf 1)=u(t)+v(t),
\qquad
Y_t(\mathbf 0)=u(t)-v(t),
\]
the integrand can be written as
\begin{align*}
&
\frac{E^1(t)}{p_G}\{u(t)+v(t)\}
-
\frac{E^0(t)}{p_G}\{u(t)-v(t)\}
-2v(t)
\\
&\qquad =
u(t)\frac{E^1(t)-E^0(t)}{p_G}
+
v(t)\left\{
\frac{E^1(t)+E^0(t)}{p_G}-2
\right\}
\\
&\qquad =
u(t)X(t)+v(t)H(t),
\end{align*}
which proves the first identity.

To prove the decomposition of the squared error~\eqref{eq:errorsqd_decomp}, we start by showing that $\widehat\tau_T$ is unbiased for $\tau_T$.
Recall
\[
\PP[G]{E^1(t)=1}
=
\PP[G]{E^0(t)=1}
=
p_G,
\]
and thus by definition,
\[
\EE[G]{X(t)}=\EE[G]{H(t)}=0.
\]
By  Assumption~\ref{assu:outcome} and Fubini's theorem, it follows from~\eqref{eq:error_decomp} that
$\EE[G]{\widehat\tau_T-\tau_T}=0$, so the estimator is unbiased.

Since the estimator is unbiased, it remains to compute its variance. Squaring the error decomposition produces two within-component terms \(X(t)X(s)\) and \(H(t)H(s)\), and a cross term \(X(t)H(s)\). We first show that the cross term is zero. Since \(X(t)\) and \(H(s)\) have mean zero,
\[
\Cov[G]{X(t),H(s)}=\EE[G]{X(t)H(s)}.
\]
By definition,
\begin{align*}
\EE[G]{X(t)H(s)}
&=
\frac{1}{p_G^2}
\EE[G]{
    \{E^1(t)-E^0(t)\}\{E^1(s)+E^0(s)\}
}
-
\frac{2}{p_G}
\EE[G]{E^1(t)-E^0(t)}.
\end{align*}
The second term is zero since
$\PP[G]{E^1(t)=1}=\PP[G]{E^0(t)=1}=p_G$. Expanding the first term gives
\begin{align*}
\EE[G]{X(t)H(s)}
=
\frac{1}{p_G^2}\Big\{&
\PP[G]{E^1(t)=1,E^1(s)=1}
+
\PP[G]{E^1(t)=1,E^0(s)=1}
\\
&-
\PP[G]{E^0(t)=1,E^1(s)=1}
-
\PP[G]{E^0(t)=1,E^0(s)=1}
\Big\}.
\end{align*}
By treatment-label symmetry,
\[
\PP[G]{E^1(t)=1,E^1(s)=1}
=
\PP[G]{E^0(t)=1,E^0(s)=1},
\]
and
\[
\PP[G]{E^1(t)=1,E^0(s)=1}
=
\PP[G]{E^0(t)=1,E^1(s)=1}.
\]
Thus, $\Cov[G]{X(t),H(s)}=0$ for every $t,s$.

Again using the error identity~\eqref{eq:error_decomp}, Assumption~\ref{assu:outcome} and Fubini's theorem,
\begin{align*}
\EE[G]{(\widehat\tau_T-\tau_T)^2}
&=
\frac{1}{T^2}
\int_0^T\int_0^T
\EE[G]{
\{u(t)X(t)+v(t)H(t)\}
\{u(s)X(s)+v(s)H(s)\}
}
\,ds\,dt.
\end{align*}
We have shown that the cross terms are all zero. Since $X$ and $H$
are stationary and mean zero,
\[
\EE[G]{X(t)X(s)}=\gamma_X(t-s),
\qquad
\EE[G]{H(t)H(s)}=\gamma_H(t-s).
\]
Therefore,
\[
\EE[G]{(\widehat\tau_T-\tau_T)^2}
=
\frac{1}{T^2}
\int_0^T\int_0^T
\left\{
u(t)u(s)\gamma_X(t-s)
+
v(t)v(s)\gamma_H(t-s)
\right\}
\,ds\,dt,
\]
which proves the second identity.

\subsection{Proof of Theorem~\ref{theo:representation}}

In Proposition~\ref{prop:decomp}, we express the mean squared error as two quadratic forms in the
autocovariance kernels. We first rewrite these quadratic forms in the frequency
domain.

By the definition of the normalized spectral density and Fourier inversion,
\[
\gamma_X(t-s)
=
\frac{1}{2\pi}
\int_{\mathbb R}
f_X(x)e^{ix(t-s)/m}\,dx.
\]
Substituting this representation into the $X$-component of Proposition~\ref{prop:decomp}
and exchanging the order of integration gives
\begin{align*}
&\int_0^T\int_0^T
u(t)u(s)\gamma_X(t-s)\,ds\,dt \\
&\qquad =
\frac{1}{2\pi}
\int_{\mathbb R}
f_X(x)
\left\{\int_0^T u(t)e^{ixt/m}\,dt\right\}
\left\{\int_0^T u(s)e^{-ixs/m}\,ds\right\}
dx \\
&\qquad =
\frac{m^2}{2\pi}
\int_{\mathbb R}
f_X(x)|U_T(x)|^2\,dx,
\end{align*}
where the last equality follows from
\[
U_T(x)
=
\frac{1}{m}\int_0^T u(t)e^{-ixt/m}\,dt.
\]
The same argument gives
\[
\int_0^T\int_0^T
v(t)v(s)\gamma_H(t-s)\,ds\,dt
=
\frac{m^2}{2\pi}
\int_{\mathbb R}
f_H(x)|V_T(x)|^2\,dx.
\]
Substituting these identities into Proposition~\ref{prop:decomp} yields
\[
\EE[G]{(\widehat\tau_T-\tau_T)^2}
=
\frac{m^2}{2\pi T^2}
\int_{\mathbb R}
\left\{
f_X(x)|U_T(x)|^2
+
f_H(x)|V_T(x)|^2
\right\}dx,
\]
which proves~\eqref{eq:representation}.

We next derive the uniform bound. Write
\[
J(G):=\sup_{x\in\mathbb R}\max\{f_X(x),f_H(x)\}.
\]
Since $f_X$ and $f_H$ are spectral densities, they are nonnegative, and hence
\begin{align*}
\EE[G]{(\widehat\tau_T-\tau_T)^2}
&\le
\frac{m^2J(G)}{2\pi T^2}
\int_{\mathbb R}
\bigl\{|U_T(x)|^2+|V_T(x)|^2\bigr\}\,dx.
\end{align*}
Using Parseval's identity,
\[
\frac{1}{2\pi}
\int_{\mathbb R}|U_T(x)|^2\,dx
=
\frac{1}{m}\int_0^T u(t)^2\,dt,
\]
and similarly for $V_T$. Therefore,
\begin{align*}
\EE[G]{(\widehat\tau_T-\tau_T)^2}
&\le
\frac{mJ(G)}{T^2}
\int_0^T\{u(t)^2+v(t)^2\}\,dt.
\end{align*}
But
\[
u(t)^2+v(t)^2
=
\frac{Y_t(\mathbf 1)^2+Y_t(\mathbf 0)^2}{2}
\le B^2
\]
under Assumption~\ref{assu:outcome}. Thus
\[
\sup_{Y\in\mathcal Y_B}
\EE[G]{(\widehat\tau_T-\tau_T)^2}
\le
\frac{B^2m}{T}J(G),
\]
which proves~\eqref{eq:proxy}.

It remains to show that the bound is asymptotically sharp when the largest
spectral value occurs at zero. Suppose first that $J(G)=f_X(0)$.
Consider the admissible constant outcome schedule
\[
Y_t(\mathbf 1)=Y_t(\mathbf 0)=B.
\]
Then $u(t)=B$ and $v(t)=0$. By Proposition~\ref{prop:decomp},
\[
\EE[G]{(\widehat\tau_T-\tau_T)^2}
=
\frac{B^2}{T^2}
\int_0^T\int_0^T
\gamma_X(t-s)\,ds\,dt.
\]
Writing the double integral in terms of the time lag gives
\[
\int_0^T\int_0^T
\gamma_X(t-s)\,ds\,dt
=
\int_{-T}^T
(T-|h|)\gamma_X(h)\,dh.
\]
Therefore,
\[
\frac{T}{B^2m}
\EE[G]{(\widehat\tau_T-\tau_T)^2}
=
\frac{1}{m}
\int_{\mathbb R}
\left(1-\frac{|h|}{T}\right)_+
\gamma_X(h)\,dh.
\]
After the change of variables $h=ms$, this becomes
\[
\int_{\mathbb R}
\left(1-\frac{m|s|}{T}\right)_+
\gamma_X(ms)\,ds.
\]
As $T/m\to\infty$, the term in parentheses converges pointwise to one
and is bounded by one. Assumption~\ref{assu:spectral} therefore allows us to apply dominated
convergence, which gives
\[
\frac{T}{B^2m}
\EE[G]{(\widehat\tau_T-\tau_T)^2}
\to
\int_{\mathbb R}\gamma_X(ms)\,ds
=
f_X(0)
=
J(G)
\]
as $T/m\to\infty$.

If instead $J(G)=f_H(0)$, choose
\[
Y_t(\mathbf 1)=B,
\qquad
Y_t(\mathbf 0)=-B.
\]
Then $u(t)=0$ and $v(t)=B$, and the same argument gives
\[
\frac{T}{B^2m}
\EE[G]{(\widehat\tau_T-\tau_T)^2}
\to
f_H(0)
=
J(G)
\]
as $T/m\to\infty$.
Thus, the supremum over $\mathcal Y_B$ has asymptotic lower bound $J(G)$.
Together with the upper bound already established, this proves the
asymptotic equality~\eqref{eq:asyp_sharp}.

\subsection{Proof of Proposition~\ref{prop:renewal_spectral}}
Because Proposition~\ref{prop:renewal_spectral} is stated on the normalized time scale, we may without loss of generality set $m=1$. To derive the stated forms of \(f_X\) and \(f_H\), we first use the following finite-window characterization of a spectral density.

\begin{lemma}
\label{lemm:long_run_fourier}
Let $Y(t)$ be a stationary mean-zero process with absolutely integrable
autocovariance function $\gamma_Y$. Then, for every $x\in\mathbb R$,
\begin{equation}
    f_Y(x)
    =
    \lim_{L\to\infty}
    \frac{1}{L}
    \EE[G]{
        \left|
            \int_0^L Y(t)e^{-ixt}\,dt
        \right|^2
    },
\end{equation}
where $f_Y(x):=\int_{\RR}\gamma_Y(h)e^{-ixh}\,dh$.
\end{lemma}

Applying Lemma~\ref{lemm:long_run_fourier} to $X$ and $H$, we need to
evaluate the resulting long-run Fourier integrals using the renewal structure. 

Recall that
\[
    X(t)=\frac{E^1(t)-E^0(t)}{p_G},
    \qquad
    H(t)=\frac{E^1(t)+E^0(t)}{p_G}-2.
\]
Thus, the Fourier integrals in Lemma~\ref{lemm:long_run_fourier} are determined
by when $E^1(t)$ and $E^0(t)$ equal one. Let $S_j$ denote successive renewal
times and let $Z_j:=S_{j+1}-S_j$ be the corresponding normalized durations.
On the run $[S_j,S_{j+1})$, both exposure indicators are zero during the first
unit of time. If $Z_j>1$, then after time $S_j+1$ the indicator corresponding
to the current treatment stays one until the run ends. 

Consider a run whose two boundaries lie inside the Fourier window,
$0\le S_j<S_{j+1}\le L$. Its contribution to
\[
    \int_0^L \{E^1(t)-E^0(t)\}e^{-ixt}\,dt
\]
is, up to the sign of the treatment assigned to that run,
\[
\begin{aligned}
    \1\{Z_j>1\}
    \int_{S_j+1}^{S_{j+1}} e^{-ixt}\,dt
    &=
    e^{-ixS_j}
    \1\{Z_j>1\}
    \int_1^{Z_j} e^{-ixs}\,ds \\
    &=
    e^{-ixS_j}\psi_{Z_j}(x).
\end{aligned}
\]
The same term contributes to
$\int_0^L \{E^1(t)+E^0(t)\}e^{-ixt}\,dt$ with a positive sign regardless of
the treatment label. 

The calculation above leaves out at most two runs: the run containing time $0$
and the run containing time $L$. However, those two runs do not affect the
long-run limit. Indeed, for any interval $[a,b]$ and fixed $x\neq0$,
\[
    \left|
        \int_a^b e^{-ixt}\,dt
    \right|
    =
    \left|
        \frac{e^{-ixa}-e^{-ixb}}{ix}
    \right|
    \le \frac{2}{|x|}.
\]
Thus, the total contribution of the two boundary runs to either Fourier
integral is bounded uniformly by $4/(p_G|x|)$. For $H$, the centering
term is also uniformly bounded, since
\[
    \left|
        2\int_0^L e^{-ixt}\,dt
    \right|
    \le \frac{4}{|x|}.
\]

These omitted terms do not affect the limit in
Lemma~\ref{lemm:long_run_fourier}. By~\eqref{eq:dct_cound},
\[
    \EE[G]{
        \left|
            \int_0^L Y(t)e^{-ixt}\,dt
        \right|^2
    }
    \le
    L\int_{\mathbb R}|\gamma_Y(h)|\,dh
    =
    O(L).
\]
Therefore, the cross-product between the full Fourier integral and any of the
uniformly bounded omitted terms is $O(\sqrt L)$ by Cauchy--Schwarz, while the
squared omitted terms are $O(1)$. After division by $L$, both vanish. Thus,
for $x\neq0$, it suffices to compute the spectral limit using only the
complete-run contributions.

Having removed the boundary and centering terms, consider $n$ consecutive
complete run-block segments and index them by $j=1,\ldots,n$. Partitioning $[0,L]$ into run segments and up to the common factor
$1/p_G$, their contribution to the Fourier integral for $H$ is
\[
    \sum_{j=1}^n e^{-ixS_j}\psi_{Z_j}(x),
\]
while for $X$ the same terms enter with alternating signs,
\[
    \sum_{j=1}^n (-1)^j e^{-ixS_j}\psi_{Z_j}(x).
\]
The initial treatment label only changes the overall sign of the latter sum and
therefore has no effect on its squared magnitude. Expanding the square
for $H$ gives
\begin{equation}
\label{eq:expansion}
\begin{aligned}
&\EE[G]{
    \left|
        \sum_{j=1}^n e^{-ixS_j}\psi_{Z_j}(x)
    \right|^2
}\\
&\qquad =
n\,\EE[G]{|\psi_Z(x)|^2}  +
2\recal\left\{
    \sum_{k=1}^{n-1}\sum_{j=1}^{n-k}
    \EE[G]{
        e^{ix(S_{j+k}-S_j)}
        \psi_{Z_j}(x)\overline{\psi_{Z_{j+k}}(x)}
    }
\right\}.
\end{aligned}
\end{equation}
For $X$, the diagonal term is exactly the same, while the cross-product between runs
$k$ apart is multiplied by $(-1)^k$.

It remains to evaluate the cross-product between two runs $k$ apart. Since
\[
    S_{j+k}-S_j
    =
    Z_j+Z_{j+1}+\cdots+Z_{j+k-1},
\]
and the durations are i.i.d., we obtain
\begin{equation}
\label{eq:characteristic}
\begin{aligned}
&\EE[G]{
    e^{ix(S_{j+k}-S_j)}
    \psi_{Z_j}(x)\overline{\psi_{Z_{j+k}}(x)}
} \\
&\qquad=\EE[G]{e^{ixZ_j}\psi_{Z_j}(x)}
    \cb{\EE[G]{e^{ixZ}}}^{k-1}
    \EE[G]{\overline{\psi_{Z_{j+k}}(x)}} \\
&\qquad=
\EE[G]{\psi_Z(x)e^{ixZ}}\,
\phi_G(x)^{k-1}\,
\EE[G]{\overline{\psi_Z(x)}},
\end{aligned}
\end{equation}
where $\phi_G(x)$ is the characteristic function of distribution $Z\sim G$.

Substituting~\eqref{eq:characteristic} into~\eqref{eq:expansion} and dividing by $n$ gives, for $H$,
\[
\begin{aligned}
&\frac{1}{n}
\EE[G]{
    \left|
        \sum_{j=1}^n e^{-ixS_j}\psi_{Z_j}(x)
    \right|^2
}\\
&\qquad=
\EE[G]{|\psi_Z(x)|^2} +
2\recal\left\{
    \EE[G]{\psi_Z(x)e^{ixZ}}
    \EE[G]{\overline{\psi_Z(x)}}
    \sum_{k=1}^{n-1}
    \left(1-\frac{k}{n}\right)\phi_G(x)^{k-1}
\right\}.
\end{aligned}
\]
For $X$, the same expression holds with an additional factor $(-1)^k$ inside
the sum. Since $G$ has a density, $|\phi_G(x)|<1$ for $x\neq0$. Hence, as
$n\to\infty$,
\[
    \sum_{k=1}^{n-1}
    \left(1-\frac{k}{n}\right)\phi_G(x)^{k-1}
    \to
    \frac{1}{1-\phi_G(x)},
\]
whereas
\[
    \sum_{k=1}^{n-1}
    \left(1-\frac{k}{n}\right)(-1)^k\phi_G(x)^{k-1}
    \to
    -\frac{1}{1+\phi_G(x)}.
\]
Substituting these limits into the expansion gives the Fourier power
per complete run. For $H$,
\begin{subequations}
\begin{equation}
\label{eq:res_H}
    \begin{aligned}
&\lim_{n\to\infty}
\frac{1}{n}
\EE[G]{
    \left|
        \sum_{j=1}^n e^{-ixS_j}\psi_{Z_j}(x)
    \right|^2
}\\
&\qquad=
\EE[G]{|\psi_Z(x)|^2} 
+
2\recal\left\{
    \frac{
        \EE[G]{\psi_Z(x)e^{ixZ}}
        \EE[G]{\overline{\psi_Z(x)}}
    }{
        1-\phi_G(x)
    }
\right\},
\end{aligned}
\end{equation}
whereas the alternating signs for $X$ give
\begin{equation}
\label{eq:res_X}
\begin{aligned}
&\lim_{n\to\infty}
\frac{1}{n}
\EE[G]{
    \left|
        \sum_{j=1}^n (-1)^j e^{-ixS_j}\psi_{Z_j}(x)
    \right|^2
}\\
&\qquad=
\EE[G]{|\psi_Z(x)|^2} 
-
2\recal\left\{
    \frac{
        \EE[G]{\psi_Z(x)e^{ixZ}}
        \EE[G]{\overline{\psi_Z(x)}}
    }{
        1+\phi_G(x)
    }
\right\}.
\end{aligned}
\end{equation}
\end{subequations}

The calculation above gives the limiting Fourier power per run, whereas Lemma~\ref{lemm:long_run_fourier} requires Fourier power per unit time. 
Denote the number of
complete runs observed over a window of length $L$ by $N_L$. We show in the following
lemma that passing from a fixed number of runs to the random number observed
over a time window simply scales the limiting Fourier power by the asymptotic
run rate $1/\EE[G]{Z}$, as is suggested by renewal theory \citep{ross1996stochastic}.
Combining Lemma~\ref{lemm:random_index} with \eqref{eq:res_H}-\eqref{eq:res_X} and restoring the factor $1/p_G^2$ from the definitions of $X$ and $H$ then gives the claimed
results for $f_X(x)$ and $f_H(x)$ when $x\neq0$.

\begin{lemma}
Under assumptions of Proposition~\ref{prop:renewal_spectral}, for any $x\neq 0$,
\begin{equation}
\lim_{L\to\infty}
\frac{1}{L}
\EE[G]{
\left|
\sum_{j=1}^{N_L}
e^{-ixS_j}\psi_{Z_j}(x)
\right|^2
}
=
\frac{1}{\EE[G]{Z}}
\lim_{n\to\infty}
\frac{1}{n}
\EE[G]{
\left|
\sum_{j=1}^{n}
e^{-ixS_j}\psi_{Z_j}(x)
\right|^2
},
\label{eq:H_target}
\end{equation}
and
\begin{equation}
\lim_{L\to\infty}
\frac{1}{L}
\EE[G]{
\left|
\sum_{j=1}^{N_L}
(-1)^j e^{-ixS_j}\psi_{Z_j}(x)
\right|^2
}
=
\frac{1}{\EE[G]{Z}}
\lim_{n\to\infty}
\frac{1}{n}
\EE[G]{
\left|
\sum_{j=1}^{n}
(-1)^j e^{-ixS_j}\psi_{Z_j}(x)
\right|^2
}.
\label{eq:X_target}
\end{equation}
\label{lemm:random_index}
\end{lemma}

It remains to evaluate the spectra at $x=0$. By
Assumption~\ref{assu:spectral}, both spectral densities are continuous at
zero. Indeed, for $Y\in\{X,H\}$,
\[
    f_Y(x)
    =
    \int_{\mathbb R}\gamma_Y(h)e^{-ixh}\,dh,
\]
and as $x\to0$, $e^{-ixh}\to1$ for every $h$, while
\[
    \left|\gamma_Y(h)e^{-ixh}\right|
    =
    |\gamma_Y(h)|.
\]
Since $\gamma_Y$ is absolutely integrable, dominated convergence gives
$f_Y(x)\to f_Y(0)$. We may therefore obtain the values at zero by taking
$x\to0$ in~\eqref{eq:res_H} and~\eqref{eq:res_X} above.

For~\eqref{eq:res_X}, note that as $x\to0$,
\[
    \psi_Z(x)\to (Z-1)_+,
    \qquad
    \phi_G(x)\to 1.
\]
Moreover, $|\psi_Z(x)|\le (Z-1)_+\le Z$, so
$\EE[G]{Z^2}<\infty$ allows another application of dominated convergence to
the expectations involving $\psi_Z(x)$. Hence
\[
\begin{aligned}
    f_X(0)
    &=
    \frac{1}{p_G^2\EE[G]{Z}}
    \left\{
        \EE[G]{(Z-1)_+^2}
        -
        \EE[G]{(Z-1)_+}^2
    \right\} \\
    &=
    \frac{\Var[G]{(Z-1)_+}}
         {p_G^2\EE[G]{Z}}.
\end{aligned}
\]

For~\eqref{eq:res_H}, we need to expand both the numerator and denominator of the cross-run term around zero. Using $\EE[G]{Z^2}<\infty$,
\[
    \phi_G(x)
    =
    1+ix\EE[G]{Z}
    -\frac{x^2}{2}\EE[G]{Z^2}
    +o(x^2),
\]
while
\[
    \EE[G]{\psi_Z(x)e^{ixZ}}
    =
    \EE[G]{(Z-1)_+}
    +
    \frac{ix}{2}\EE[G]{(Z-1)_+^2}
    +o(x),
\]
and
\[
    \EE[G]{\overline{\psi_Z(x)}}
    =
    \EE[G]{(Z-1)_+}
    +
    \frac{ix}{2}
    \EE[G]{(Z^2-1)\1\{Z>1\}}
    +o(x).
\]
Substituting these expansions into the expression for $f_H(x)$ and using
\[
    \EE[G]{(Z-1)_+}
    =
    2p_G\EE[G]{Z},
\]
together with
\[
    (Z-1)_+^2
    +(Z^2-1)\1\{Z>1\}
    =
    2Z(Z-1)_+,
\]
gives
\[
    f_H(0)
    =
    \frac{
        \Var[G]{(Z-1)_+-2p_GZ}
    }{
        p_G^2\EE[G]{Z}
    }.
\]
This completes the proof.

\subsection{Proof of Corollary~\ref{coro:regular}}

Because the spectra are defined on the normalized time scale, we may without loss of generality set
$m=\ell=1$.  Let $\varepsilon_k\in\{-1,1\}$ denote the treatment label of
block $k$, where $\{\varepsilon_k\}_{k\in\mathbb Z}$ are i.i.d. Rademacher
random variables. We randomize the block origin uniformly over $[0,1)$
to make the process stationary.

For almost every time $t$ in block $k$, the preceding unit interval
$[t-1,t]$ overlaps blocks $k-1$ and $k$.  Thus $t$ is under pure exposure
if and only if $\varepsilon_{k-1}=\varepsilon_k$, and hence
$p_{G_{\rm reg}}=1/4$.  It follows that $X(t)$ and $H(t)$ are constant
within each block, with block-level values
\[
X_k
    = 2(\varepsilon_{k-1}+\varepsilon_k),
    \qquad
H_k
    = 2\varepsilon_{k-1}\varepsilon_k.
\]
Therefore,
\[
\EE[G_{\rm reg}]{X_k^2}=8,
\qquad
\EE[G_{\rm reg}]{X_kX_{k+1}}=4,
\qquad
\EE[G_{\rm reg}]{X_kX_{k+j}}=0
\quad\text{for } |j|\ge 2,
\]
while
\[
\EE[G_{\rm reg}]{H_k^2}=4,
\qquad
\EE[G_{\rm reg}]{H_kH_{k+j}}=0
\quad\text{for } j\neq 0.
\]

The (uniformly) random block origin makes the process stationary and allows us to
express its continuous-time autocovariances in terms of the block-level
covariances. For $0\le h\le 1$, two times separated by $h$ fall in the
same block with probability $1-h$ and in adjacent blocks with probability
$h$. Hence
\[
\gamma_X(h)
=8(1-h)+4h
=4(2-h),
\qquad
\gamma_H(h)
=4(1-h).
\]
For $1\le h\le 2$, the two times are in adjacent blocks with probability
$2-h$ and otherwise are at least two blocks apart. Therefore,
\[
\gamma_X(h)=4(2-h),
\qquad
\gamma_H(h)=0.
\]
For $h\ge 2$, both autocovariances are zero. By symmetry,
\[
\gamma_X(h)=4(2-|h|)_+,
\qquad
\gamma_H(h)=4(1-|h|)_+.
\]

Taking Fourier transforms of these triangular functions gives
\[
f_X(x)
    =16\left(\frac{\sin x}{x}\right)^2,
\qquad
f_H(x)
    =4\left(\frac{\sin(x/2)}{x/2}\right)^2.
\]
At zero,
\[
\lim_{x\to 0}\frac{\sin x}{x}
=
\lim_{x\to 0}\frac{\sin(x/2)}{x/2}
=
1.
\]  
Since
$|\sin y|\le |y|$,
\[
\sup_{x\in\mathbb R}\max\{f_X(x),f_H(x)\}
    = f_X(0)=16.
\]
Restoring the original time scale and applying Theorem~2 therefore yields
\[
\sup_{Y\in\mathcal Y_B}
\EE[G_{\rm reg}]{(\widehat\tau_T-\tau_T)^2}
\le \frac{16B^2m}{T}.
\]
Moreover, the spectral supremum is attained at zero frequency, so the
asymptotic sharpness statement in Theorem~2 gives
\[
\lim_{T/m\to\infty}
\frac{T}{B^2m}
\sup_{Y\in\mathcal Y_B}
\EE[G_{\rm reg}]{(\widehat\tau_T-\tau_T)^2}
=16.
\]

\subsection{Proof of Proposition~\ref{prop:l2_sharpness}}

Under Assumption~\ref{assu:l2_avg}, Proposition~\ref{prop:decomp} and the spectral representation~\eqref{eq:representation} continue to hold, since the second-moment condition is sufficient for the Fubini and Parseval arguments in which Assumption~\ref{assu:outcome} was used. We therefore take these results as given here.

The upper bound follows from the same argument as that of
Theorem~\ref{theo:representation}. Indeed, since
\[
u(t)^2+v(t)^2
=
\frac{Y_t(\mathbf 1)^2+Y_t(\mathbf 0)^2}{2},
\]
for any $Y\in\mathcal Y_{B,2}(T)$ (i.e., the constrained second-moment class),
\[
\int_0^T\{u(t)^2+v(t)^2\}\,dt
\le B^2T.
\]
Applying the spectral representation and Parseval's identity therefore gives
\[
\sup_{Y\in\mathcal Y_{B,2}(T)}
\EE[G]{(\widehat\tau_T-\tau_T)^2}
\le
\frac{B^2m}{T}J(G).
\]

It remains to show that this upper bound can be attained asymptotically.
Under Assumption~\ref{assu:spectral}, the autocovariance functions
$\gamma_X$ and $\gamma_H$ are absolutely integrable. Their Fourier
transforms $f_X$ and $f_H$ are therefore continuous and satisfy
\[
f_X(x)\to 0,
\qquad
f_H(x)\to 0
\qquad\text{as } |x|\to\infty,
\]
by the Riemann-Lebesgue lemma. As a result,
$\max\{f_X(x),f_H(x)\}$ is continuous and tends to zero as
$|x|\to\infty$, so its supremum is attained at some frequency
$x^\star$. 

Since both spectra are symmetric around zero, we may take
$x^\star\ge 0$. The case
$x^\star=0$ is attained by the constant outcome schedules used in the
proof of Theorem~\ref{theo:representation}, so suppose $x^\star>0$.

First suppose that
\[
J(G)=f_X(x^\star).
\]
To attain this value, we would like the outcome variation entering the
$X$-component of Proposition~\ref{prop:decomp} to oscillate at exactly the frequency
$x^\star$. We therefore choose the outcomes under global treatment and global control to be the same sinusoid. Define
\[
c_T^2
:=
\frac{1}{T}\int_0^T
\cos^2\left(\frac{x^\star t}{m}\right)\,dt
\]
and set
\[
Y_t(\mathbf 1)
=
Y_t(\mathbf 0)
=
\frac{B}{c_T}
\cos\left(\frac{x^\star t}{m}\right).
\]
Then
\[
u(t)
=
\frac{B}{c_T}
\cos\left(\frac{x^\star t}{m}\right),
\qquad
v(t)=0,
\]
and the second-moment constraint is satisfied with equality. Moreover,
\[
c_T^2
=
\frac12
+
\frac{m}{4x^\star T}
\sin\left(\frac{2x^\star T}{m}\right)
\to
\frac12
\]
as $T/m\to\infty$.

We now show that this outcome schedule makes the normalized mean squared
error converge to $f_X(x^\star)$. By Proposition~\ref{prop:decomp},
\begin{align*}
\frac{T}{B^2m}
\EE[G]{(\widehat\tau_T-\tau_T)^2}
&=
\frac{1}{m c_T^2}
\int_{-T}^T
\gamma_X(h)
\left\{
\frac{1}{T}
\int_{[0,T]\cap[h,T+h]}
\cos\left(\frac{x^\star t}{m}\right)
\cos\left(\frac{x^\star(t-h)}{m}\right)
\,dt
\right\}dh.
\end{align*}
Here we have grouped pairs of time points according to their lag
$h=t-s$. For a fixed lag $h$, the two time points $t$ and $t-h$ must both lie in $[0,T]$; this is why we only take integral for
$t\in[0,T]\cap[h,T+h]$.

It therefore remains to understand the long-run average product of two
copies of the sinusoid separated by a fixed lag $h$. Using the
product-to-sum identity,
\begin{align*}
&\frac{1}{T}
\int_{[0,T]\cap[h,T+h]}
\cos\left(\frac{x^\star t}{m}\right)
\cos\left(\frac{x^\star(t-h)}{m}\right)
\,dt \\
&\qquad =
\frac{T-|h|}{2T}
\cos\left(\frac{x^\star h}{m}\right)
+
\frac{1}{2T}
\int_{[0,T]\cap[h,T+h]}
\cos\left(\frac{2x^\star t-x^\star h}{m}\right)
\,dt
\end{align*}
for $T>|h|$. The first term converges to
\[
\frac12
\cos\left(\frac{x^\star h}{m}\right),
\]
while the integral in the second term remains bounded as $T$ grows, so
the second term converges to zero. Thus, for every fixed $h$,
\[
\frac{1}{c_T^2 T}
\int_{[0,T]\cap[h,T+h]}
\cos\left(\frac{x^\star t}{m}\right)
\cos\left(\frac{x^\star(t-h)}{m}\right)
\,dt
\to
\cos\left(\frac{x^\star h}{m}\right).
\]

Since $c_T^2\to1/2$ and the cosine terms are bounded by one, 
\begin{align*}
\frac{1}{Tc_T^2}
\int_{[0,T]\cap[h,T+h]}
\cos\left(\frac{x^\star t}{m}\right)
\cos\left(\frac{x^\star(t-h)}{m}\right)
\,dt
\end{align*}
is uniformly
bounded in $h$ for all sufficiently large $T$. Assumption~\ref{assu:spectral} therefore allows us
to apply dominated convergence, which gives
\begin{align*}
\frac{T}{B^2m}
\EE[G]{(\widehat\tau_T-\tau_T)^2}
&\to
\frac{1}{m}
\int_{\mathbb R}
\gamma_X(h)
\cos\left(\frac{x^\star h}{m}\right)\,dh \\
&=
f_X(x^\star)
=
J(G).
\end{align*}
The second equality follows from the definition of the normalized spectral
density and the symmetry of the autocovariance function.

If instead
\[
J(G)=f_H(x^\star),
\]
we use the same sinusoid with opposite signs under global treatment and
control:
\[
Y_t(\mathbf 1)
=
\frac{B}{c_T}
\cos\left(\frac{x^\star t}{m}\right),
\qquad
Y_t(\mathbf 0)
=
-\frac{B}{c_T}
\cos\left(\frac{x^\star t}{m}\right).
\]
Then $u(t)=0$ and
\[
v(t)
=
\frac{B}{c_T}
\cos\left(\frac{x^\star t}{m}\right),
\]
so the same argument yields
\[
\frac{T}{B^2m}
\EE[G]{(\widehat\tau_T-\tau_T)^2}
\to
f_H(x^\star)
=
J(G).
\]
Thus, the spectral upper bound is asymptotically attainable whenever its
maximum is attained. Together with the upper bound above, this proves the
claim.

\subsection{Proof of Lemma~\ref{lemm:long_run_fourier}}

For any $L>0$, expanding the square and using stationarity gives
\[
\begin{aligned}
\frac{1}{L}
\EE[G]{
    \left|
        \int_0^L Y(t)e^{-ixt}\,dt
    \right|^2
}
&=
\frac{1}{L}
\int_0^L\int_0^L
    \gamma_Y(t-s)e^{-ix(t-s)}
\,dt\,ds \\
&=
\int_{-L}^L
    \left(1-\frac{|h|}{L}\right)
    \gamma_Y(h)e^{-ixh}
\,dh\\
&=
\int_{\RR}
    \left(1-\frac{|h|}{L}\right)_+
    \gamma_Y(h)e^{-ixh}
\,dh.
\end{aligned}
\]
For each fixed $h$,
\[
    \left(1-\frac{|h|}{L}\right)_+ \to 1\qquad \text{ as }L\to\infty.
\]
Furthermore, Euler's formula gives $\abs{e^{-ixh}}=1$, and thus
\begin{equation}
\label{eq:dct_cound}
    \left|
        \left(1-\frac{|h|}{L}\right)_+
        \gamma_Y(h)e^{-ixh}
    \right|
    \le |\gamma_Y(h)|.
\end{equation}
Since $\gamma_Y$ is absolutely integrable, dominated convergence yields
\[
    \lim_{L\to\infty}
    \frac{1}{L}
    \EE[G]{
        \left|
            \int_0^L Y(t)e^{-ixt}\,dt
        \right|^2
    }
    =
    \int_{\mathbb R}\gamma_Y(h)e^{-ixh}\,dh
    =
    f_Y(x).
\]

\subsection{Proof of Lemma~\ref{lemm:random_index}}

To start with, let's consider the equality for $H$ in~\eqref{eq:H_target}.
We first evaluate the limit on the right-hand side of~\eqref{eq:H_target}. Define
\[
b_G
:=
\frac{\EE[G]{\psi_Z(x)}}{1-\phi_G(-x)}.
\]
Since $\phi_G(-x)=\overline{\phi_G(x)}$ and
$\abs{\phi_G(x)}<1$ for $x\neq0$, the denominator is nonzero. By the
definition of $b_G$,
\begin{equation}
\label{eq:H_centered_mean}
\EE[G]{
\psi_Z(x)-b_G\{1-e^{-ixZ}\}
}
=
\EE[G]{\psi_Z(x)}
-
b_G\{1-\phi_G(-x)\}
=
0.
\end{equation}

Since $S_{j+1}=S_j+Z_j$,
\[
\begin{split}
e^{-ixS_j}\psi_{Z_j}(x)
={}&
e^{-ixS_j}
\left[
\psi_{Z_j}(x)-b_G\{1-e^{-ixZ_j}\}
\right] \\
&\qquad+
b_G\{e^{-ixS_j}-e^{-ixS_{j+1}}\}.
\end{split}
\]
Summing over $j=1,\ldots,n$ gives
\begin{equation}
\label{eq:H_fixed_decomposition}
\begin{split}
\sum_{j=1}^n e^{-ixS_j}\psi_{Z_j}(x)
={}&
\sum_{j=1}^n e^{-ixS_j}
\left[
\psi_{Z_j}(x)-b_G\{1-e^{-ixZ_j}\}
\right] \\
&\qquad+
b_G\{e^{-ixS_1}-e^{-ixS_{n+1}}\}.
\end{split}
\end{equation}
To calculate the expected square of the first sum in
\eqref{eq:H_fixed_decomposition}, we condition on
$S_1,Z_1,\ldots,Z_{k-1}$ for $j<k$. Everything involving run $j$, as well as
$S_k$, is then fixed, while $Z_k$ is an independent draw from $G$.
Thus, by \eqref{eq:H_centered_mean},
\begin{equation}
\label{eq:H_cross_zero}
\begin{split}
&\EE[G]{
e^{-ixS_j}
\left[
\psi_{Z_j}(x)-b_G\{1-e^{-ixZ_j}\}
\right]
\overline{
e^{-ixS_k}
\left[
\psi_{Z_k}(x)-b_G\{1-e^{-ixZ_k}\}
\right]
}
} \\
&=
\EE[G]{
e^{-ixS_j}
\left[
\psi_{Z_j}(x)-b_G\{1-e^{-ixZ_j}\}
\right]
e^{ixS_k}
\EE[G]{
\overline{
\psi_{Z_k}(x)-b_G\{1-e^{-ixZ_k}\}
}
\mid
S_1,Z_1,\ldots,Z_{k-1}
}
} \\
&=0.
\end{split}
\end{equation}
Thus, all cross-products between different runs vanish. Since
$\abs{e^{-ixS_j}}=1$,
\begin{equation}
\label{eq:H_centered_square}
\begin{split}
&\EE[G]{
\left|
\sum_{j=1}^n
e^{-ixS_j}
\left[
\psi_{Z_j}(x)-b_G\{1-e^{-ixZ_j}\}
\right]
\right|^2
} \\
&\qquad=
\sum_{j=1}^n
\EE[G]{
\left|
\psi_{Z_j}(x)-b_G\{1-e^{-ixZ_j}\}
\right|^2
} \\
&\qquad=
n\,
\EE[G]{
\left|
\psi_Z(x)-b_G\{1-e^{-ixZ}\}
\right|^2
}.
\end{split}
\end{equation}
Returning to \eqref{eq:H_fixed_decomposition}, note that
\[
\abs{
b_G\{e^{-ixS_1}-e^{-ixS_{n+1}}\}
}
\le 2\abs{b_G}.
\]
Therefore, using \eqref{eq:H_centered_square} and Cauchy--Schwarz,
\begin{equation}
\label{eq:H_fixed_limit}
\begin{split}
\frac{1}{n}
\EE[G]{
\left|
\sum_{j=1}^n e^{-ixS_j}\psi_{Z_j}(x)
\right|^2
}
&=
\frac{1}{n}
\EE[G]{
\left|
\sum_{j=1}^n
e^{-ixS_j}
\left[
\psi_{Z_j}(x)-b_G\{1-e^{-ixZ_j}\}
\right]
\right|^2
}
+o(1) \\
&=
\EE[G]{
\left|
\psi_Z(x)-b_G\{1-e^{-ixZ}\}
\right|^2
}
+o(1).
\end{split}
\end{equation}
As a result,
\begin{equation}
\label{eq:H_fixed_limit_final}
\lim_{n\to\infty}
\frac{1}{n}
\EE[G]{
\left|
\sum_{j=1}^n e^{-ixS_j}\psi_{Z_j}(x)
\right|^2
}
=
\EE[G]{
\left|
\psi_Z(x)-b_G\{1-e^{-ixZ}\}
\right|^2
}.
\end{equation}

We now evaluate the limit on the left-hand side of
\eqref{eq:H_target}. We first sum over all runs whose left endpoints
satisfy $S_j<L$. 
Because $Z_j$ is independent of
$S_1,Z_1,\ldots,Z_{j-1}$ and has distribution $G$,
\begin{equation}
\label{eq:H_window_mean_zero}
\begin{split}
&\EE[G]{
\1\{S_j<L\}
\left[
\psi_{Z_j}(x)-b_G\{1-e^{-ixZ_j}\}
\right]
\,\middle|\,
S_1,Z_1,\ldots,Z_{j-1}
} \\
&\qquad=
\1\{S_j<L\}
\EE[G]{
\psi_{Z_j}(x)-b_G\{1-e^{-ixZ_j}\}
}
=0.
\end{split}
\end{equation}
Thus, the cross-products between different runs again vanish. Similar to~\eqref{eq:H_centered_square}, expanding the square again gives
\begin{equation}
\label{eq:H_window_centered_square}
\begin{split}
&\EE[G]{
\left|
\sum_{j:S_j<L}
e^{-ixS_j}
\left[
\psi_{Z_j}(x)-b_G\{1-e^{-ixZ_j}\}
\right]
\right|^2
} \\
&\qquad=
\EE[G]{
\sum_{j:S_j<L}1
}
\EE[G]{
\left|
\psi_Z(x)-b_G\{1-e^{-ixZ}\}
\right|^2
}.
\end{split}
\end{equation}
By the renewal theorem~\citep{ross1996stochastic},
\begin{equation}
\label{eq:H_renewal_rate}
\frac{1}{L}
\EE[G]{
\sum_{j:S_j<L}1
}
\to
\frac{1}{\EE[G]{Z}}.
\end{equation}
Combining \eqref{eq:H_window_centered_square} and
\eqref{eq:H_renewal_rate}, as $L\to\infty$,
\begin{equation}
\label{eq:H_window_limit}
\begin{split}
&\frac{1}{L}
\EE[G]{
\left|
\sum_{j:S_j<L}
e^{-ixS_j}\psi_{Z_j}(x)
\right|^2
} \\
&=
\frac{1}{L}
\EE[G]{
\left|
\sum_{j:S_j<L}
e^{-ixS_j}
\left[
\psi_{Z_j}(x)-b_G\{1-e^{-ixZ_j}\}
\right]
+
b_G\left\{
e^{-ixS_1}
-
e^{-ixS_{\max\{j:S_j<L\}+1}}
\right\}
\right|^2
} \\
&=
\frac{1}{L}
\EE[G]{
\sum_{j:S_j<L}1
}
\EE[G]{
\left|
\psi_Z(x)-b_G\{1-e^{-ixZ}\}
\right|^2
}
+o(1) \\
&\to \frac{1}{\EE[G]{Z}}
\EE[G]{
\left|
\psi_Z(x)-b_G\{1-e^{-ixZ}\}
\right|^2
}.
\end{split}
\end{equation}

Note that the sum in \eqref{eq:H_window_limit} and the sum over the $N_L$
complete runs differ by at most the single run that starts before $L$
but ends after $L$. Since $\abs{\psi_Z(x)}
\le 2/\abs{x}$,
adding or removing this one term does not affect the limit after
division by $L$. Therefore,
\begin{equation}
\label{eq:H_complete_run_limit}
\begin{split}
&\lim_{L\to\infty}
\frac{1}{L}
\EE[G]{
\left|
\sum_{j=1}^{N_L}
e^{-ixS_j}\psi_{Z_j}(x)
\right|^2
} =
\frac{1}{\EE[G]{Z}}
\EE[G]{
\left|
\psi_Z(x)-b_G\{1-e^{-ixZ}\}
\right|^2
}.
\end{split}
\end{equation}
Substituting \eqref{eq:H_fixed_limit_final} into
\eqref{eq:H_complete_run_limit} gives the desired result~\eqref{eq:H_target}.

We now prove \eqref{eq:X_target}. For the alternating sum, redefine
\[
b_G
:=
\frac{\EE[G]{\psi_Z(x)}}{1+\phi_G(-x)}.
\]
Then
\begin{equation}
\label{eq:X_centered_mean}
\EE[G]{
\psi_Z(x)-b_G\{1+e^{-ixZ}\}
}
=
\EE[G]{\psi_Z(x)}
-
b_G\{1+\phi_G(-x)\}
=
0.
\end{equation}
Moreover, since $S_{j+1}=S_j+Z_j$,
\begin{equation}
\label{eq:X_decomposition}
\begin{split}
(-1)^j e^{-ixS_j}\psi_{Z_j}(x)
={}&
(-1)^j e^{-ixS_j}
\left[
\psi_{Z_j}(x)-b_G\{1+e^{-ixZ_j}\}
\right] \\
&+
b_G\left\{
(-1)^j e^{-ixS_j}
-
(-1)^{j+1}e^{-ixS_{j+1}}
\right\}.
\end{split}
\end{equation}
Thus, when summed over consecutive runs, the second term in
\eqref{eq:X_decomposition} telescopes and has absolute value at most
$2\abs{b_G}$.

The rest of the argument is the same as for $H$. By
\eqref{eq:X_centered_mean}, the cross-products between different centered
run contributions vanish, so for fixed $n$,
\[
\lim_{n\to\infty}
\frac{1}{n}
\EE[G]{
\left|
\sum_{j=1}^n
(-1)^j e^{-ixS_j}\psi_{Z_j}(x)
\right|^2
}
=
\EE[G]{
\left|
\psi_Z(x)-b_G\{1+e^{-ixZ}\}
\right|^2
}.
\]
Likewise, summing first over runs with $S_j<L$, the same conditional
mean-zero argument applies because $\{S_j<L\}$ is determined before
$Z_j$ is drawn, so again by the elementary renewal theorem,
\[
\lim_{L\to\infty}
\frac{1}{L}
\EE[G]{
\left|
\sum_{j:S_j<L}
(-1)^j e^{-ixS_j}\psi_{Z_j}(x)
\right|^2
}
=
\frac{1}{\EE[G]{Z}}
\EE[G]{
\left|
\psi_Z(x)-b_G\{1+e^{-ixZ}\}
\right|^2
}.
\]
Finally, replacing the runs with $S_j<L$ by the $N_L$ complete runs
changes the sum by at most one bounded term. Therefore,
\[
\lim_{L\to\infty}
\frac{1}{L}
\EE[G]{
\left|
\sum_{j=1}^{N_L}
(-1)^j e^{-ixS_j}\psi_{Z_j}(x)
\right|^2
}
=
\frac{1}{\EE[G]{Z}}
\lim_{n\to\infty}
\frac{1}{n}
\EE[G]{
\left|
\sum_{j=1}^{n}
(-1)^j e^{-ixS_j}\psi_{Z_j}(x)
\right|^2
},
\]
which proves \eqref{eq:X_target}.

\end{document}